\documentclass[aps,prl,floats,twocolumn,superscriptaddress,nobibnotes]{revtex4-2}

\usepackage[svgnames]{xcolor} 
\usepackage{graphicx,epsfig}
\usepackage{times} 
\usepackage{amssymb,amsmath,multirow,rotate,color}
\usepackage{xcolor}
\usepackage[normalem]{ulem}
\usepackage{tikz}
\usetikzlibrary{arrows.meta, positioning}

\usepackage{todonotes}
\usepackage[colorlinks=true,
						linkcolor=blue,
						urlcolor=blue,
						citecolor=blue,
						bookmarks=true,
						pdfborder={0 0 0}]{hyperref}

\definecolor{albicocca}{rgb}{0.98, 0.7, 0.2}
\definecolor{internationalorange}{rgb}{1.0, 0.31, 0.0}
\definecolor{giocolor}{RGB}{0, 150, 100}

\usepackage{float}
\usepackage{lipsum} 
\usepackage{ragged2e}
\usepackage{soul}
\DeclareMathAlphabet\mathbfcal{OMS}{cmsy}{b}{n}

\begin{document}

\title{Nested hyperedges promote the onset of collective transitions but suppress explosive behavior}

\author{Federico Malizia}
\affiliation{Department of Network and Data Science, Central European University, Vienna, Austria}
\author{Andr\'es Guzm\'an}
\affiliation{Network Science Institute, Northeastern University London, London E1W 1LP, United Kingdom}
\author{Federico Battiston}
\thanks{These authors jointly supervised this work}
\affiliation{Department of Network and Data Science, Central European University, Vienna, Austria}
\affiliation{Department of AI, Data and Decision Sciences, Luiss University of Rome, Viale Romania, Rome, Italy
}
\author{Istv\'an Z. Kiss}
\thanks{These authors jointly supervised this work}
\affiliation{Network Science Institute, Northeastern University London, London E1W 1LP, United Kingdom}
\affiliation{Department of Mathematics, Northeastern University, Boston, MA 02115, USA}
\date{\today}

\begin{abstract}

    Higher-order interactions can induce abrupt collective transitions, yet the microscopic mechanisms controlling macroscopic critical behavior remain unclear. Here we show that nested hyperedges generate a dual effect on dynamical processes: they promote the onset of collective behavior while suppressing the explosive transitions driven by higher-order feedback. To uncover the mechanism, we develop an analytically tractable theory of contagion on hypergraphs that explicitly tracks nestedness between groups of different sizes, allowing us to identify the microscopic mechanism responsible for this dual behavior. By disentangling contagion pathways, we find that nestedness redirects transmission from external links to internal, group-embedded routes---boosting early activation but making dyadic and triadic channels increasingly redundant. This loss of structural independence quenches the nonlinear amplification required for bistability, progressively smoothing the transition as hyperedges become nested. 
    The phenomenology holds for groups of any size, and is not specific to spreading dynamics but also emerges in higher-order Ising and Kuramoto dynamics. Overall, our results identify nestedness between group interactions as a general structural mechanism governing critical transitions in complex systems.
\end{abstract}
\maketitle

\textit{Introduction---}
Interactions involving groups of units, beyond pairwise ones, enrich the collective behavior of complex systems, giving rise to multistability, hysteresis, and discontinuous transitions
\cite{battiston2020networks,battiston2021physics,iacopini2019simplicial,ferraz2023multistability,skardal2020higher,civilini2024explosive,robiglio2025higher,kuehn2021universal,perez2025social}. Recent studies have shown that these phenomena depend not only on the presence of such higher-order interactions, but also on their microscopic organization~\cite{zhang2023higher,lamata2025hyperedge,burgio2024triadic,kim2023contagion,kim2024higher,malizia2025disentangling,landry2020effect,malizia2025hyperedge,keating2025loops}. 
A particularly important structural property is \emph{nestedness}, namely the tendency of smaller interactions to be embedded within larger ones. Such nested structures have been repeatedly observed in empirical systems~\cite{lotito2022higher,landry2024simpliciality,malizia2025disentangling,kirkley2025structural,larock2026nestedness}, revealing that interactions of different sizes are often organized hierarchically rather than independently. Although nestedness is known to facilitate activation in higher-order contagion processes~\cite{malizia2025disentangling,burgio2024triadic}, whether and how it controls the nature of collective transitions, in particular the emergence of backward bifurcations, leading to bistability and hysteresis, remains unknown.

Here, we identify the microscopic mechanism by which nested hyperedges simultaneously promote collective activation and suppress explosive transitions. To uncover this mechanism analytically, we use higher-order contagion as a minimal yet tractable setting. We consider a homogeneous mean-field description that allows us to characterize the emergence of backward bifurcations while explicitly disentangling external and group-embedded transmission pathways. This, in turn, provides analytical access to the early-time dynamical correlations governing the onset of collective behavior through a fast-variable approach. We show that increasing nestedness promotes early activation but progressively reduces the structural independence between pairwise and higher-order transmission, weakening the nonlinear reinforcement responsible for bistability. As a consequence, nestedness simultaneously lowers the onset threshold while shrinking the bistable region. Finally, we show that this mechanism is neither specific to contagion dynamics nor restricted to pairwise--three-body interactions. The same phenomenology emerges across different interaction orders and in higher-order Ising \cite{robiglio2025higher,son2026phase} and Kuramoto dynamics \cite{skardal2020higher}, identifying nestedness between interaction orders as a general structural principle governing both the onset and the nature of collective transitions. 

\begin{figure}[t!]
    \centering
    \includegraphics[width=\linewidth]{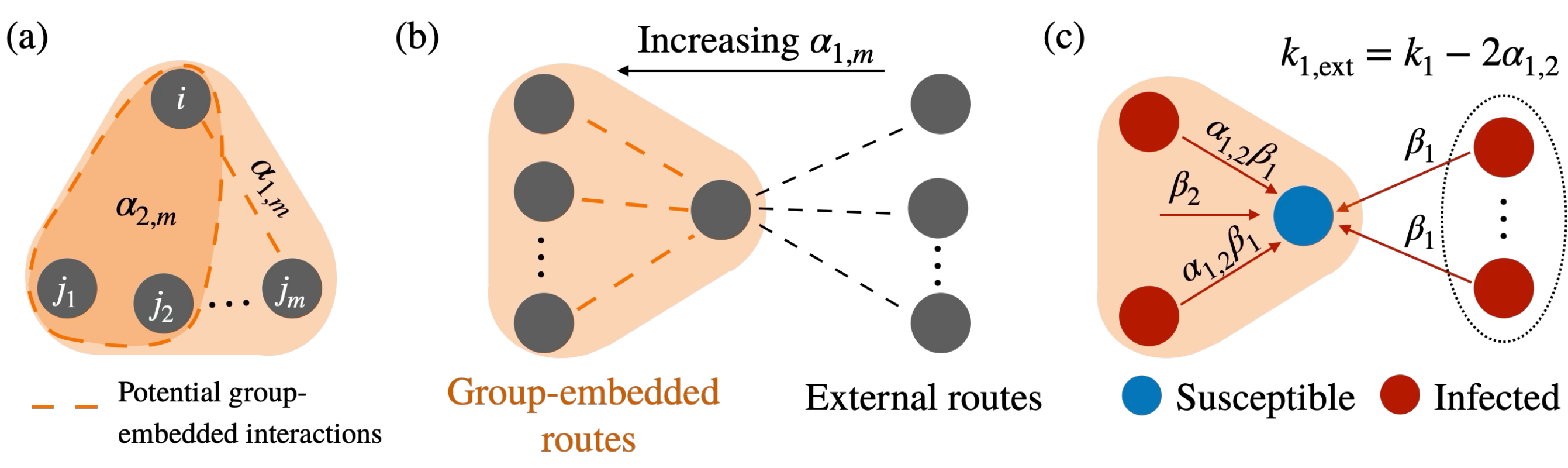}
\caption{\textbf{Nestedness across interaction orders.}
(a) Inter-order overlap $\alpha_{p,m}$ quantifies the fraction of $p$-faces of higher-order groups that are realized as lower-order interactions. Dashed connections denote potential embedded interactions. (b) Increasing nestedness redistributes lower-order interaction routes from external neighborhoods to group-embedded ones. (c) In the SIS model, this structural reorganization is incorporated analytically by decomposing pairwise transmission into embedded and external routes. When the susceptible node is conditioned on belonging to a 2-hyperedge, nestedness reduces the number of available external routes to $k_{1,\mathrm{ext}}=k_1-2\alpha_{1,2}$.}
    \label{fig:fig1}
\end{figure}

\textit{Modeling nestedness of group interactions.---}
We consider hypergraphs $\mathcal{H}=(\mathcal{N},\mathcal{E})$, where each hyperedge $e\in\mathcal{E}$ has order $m=|e|-1$, with $m=1$ denoting pairwise interactions, $m=2$ three-body interactions, and so forth. To quantify structural correlations across interaction orders, we use the \textit{inter-order hyperedge overlap}~\cite{lamata2025hyperedge}, $\alpha_{p,m}=|\mathcal{E}_{p}\cap\mathcal{F}(\mathcal{E}_{m})|/|\mathcal{F}(\mathcal{E}_{m})|$, with $p<m$, where $\mathcal{F}(\mathcal{E}_{m})$ is the set of $p$-cliques contained within the $m$-hyperedges. This quantity measures the fraction of $p$-body interactions embedded within $m$-body groups. For example, $\alpha_{1,2}=0$ corresponds to independent pairwise and three-body interactions, whereas $\alpha_{1,2}=1$ denotes complete nesting. 
Although $\alpha_{p,m}$ is defined as a global structural quantity, in homogeneous hypergraphs it can be interpreted as the probability that a randomly selected $p$-face of an $m$-hyperedge is realized as a $p$-hyperedge [Fig.~\ref{fig:fig1}(a)]. Increasing nestedness redistributes lower-order interactions from external neighborhoods to group-embedded ones. For instance, larger $\alpha_{1,m}$ reduces the number of external pairwise routes available within an $m$-body interaction while increasing the fraction of embedded ones [Fig.~\ref{fig:fig1}(b)]. Although illustrated for pairwise interactions, the same route redistribution applies to arbitrary orders $p<m$. Thus, nestedness tunes the structural independence between interaction orders, raising the question of how this reorganization shapes both the onset and the nature of collective transitions.

\textit{An analytically tractable framework.---}
To uncover the microscopic mechanism analytically, we use SIS dynamics as a minimal yet tractable setting. We consider regular hypergraphs with pairwise ($m=1$) and three-body ($m=2$) interactions, where each node belongs to exactly $k_1$ links and $k_2$ three-body groups. Susceptible nodes become infected through pairwise interactions at rate $\beta_1$ or through three-body interactions at rate $\beta_2$ when both other nodes in the group are infected, while infected nodes recover at rate $\mu$. The two transmission channels represent distinct dyadic and group-mediated mechanisms, as commonly assumed in models of social contagion~\cite{centola2007complex,iacopini2019simplicial,ferraz2023multistability}, rather than a single transmission process whose rate depends on group size.

Our homogeneous mean-field framework explicitly retains the cross-order correlations induced by nestedness by tracking node, pair, and group motifs. The density of infected nodes evolves as
\begin{equation}
\label{eq:evolution_infecteds}
\dot{\rho}^{\rm I}
=
-\mu\rho^{\rm I}
+
\beta_1 k_1\rho^{\rm SI}
+
\beta_2 k_2\rho^{\rm ISI_\Delta},
\end{equation}
where $\rho^{\rm SI}$ and $\rho^{\rm ISI_\Delta}$ denote the densities of infected--susceptible links and mixed three-body groups, respectively. Together with the evolution of the remaining motif densities, Eq.\eqref{eq:evolution_infecteds} forms a closed dynamical system. Nestedness enters the dynamics through this probabilistic interpretation: pairwise transmission is statistically decomposed into group-embedded and external routes. For a node already belonging to a three-body interaction, increasing $\alpha_{1,2}$ reduces the number of external routes in favor of internal ones, yielding $k_{1,\mathrm{ext}}=k_1-2\alpha_{1,2}$ [Fig.~\ref{fig:fig1}(c)], the only structural ingredient through which nestedness modifies the equations.
The resulting motif hierarchy is then closed using standard homogeneous approximations~\cite{kiss2017mathematics,house2009motif}, while preserving the cross-order correlations generated by nestedness. Details of the derivation are reported in the Appendix.

\textit{Nestedness reshapes critical transitions.---}
We now quantify, within this analytically tractable setting, how route redistribution induced by nestedness shapes the onset and nature of collective transitions. We use the rescaled infectivities $\lambda_1 = k_1\beta_1/\mu$ and $\lambda_2 = k_2\beta_2/\mu$, and denote by $\lambda_1^*$ the transcritical threshold and by $\hat{\lambda}_2$ the minimum group infectivity required for a backward bifurcation (bistability).

We linearize around the disease-free equilibrium $\mathbf{x}^*=(0,0,1,0,0)$ and obtain the epidemic threshold $\lambda_1^*$ analytically as the root of a quadratic equation from the linear stability condition (Appendix A). Its dependence on model parameters is not transparent in closed form; to extract the leading structural dependence, we perform an asymptotic expansion for small~$\alpha$, obtaining
\begin{equation}
\label{eq:transcritical_approximated}
\lambda_1^* \approx \frac{k_1}{k_1-1} - \alpha_{1,2}\lambda_2\frac{k_1^2}{(k_1-1)^3}.
\end{equation}
This expression shows that nestedness anticipates the epidemic onset through the combined control $\alpha_{1,2}\lambda_2$, while recovering the standard SIS threshold on networks, $\lambda_1^{*,(0)} = k_1/(k_1-1)$, in the non-nested limit $\alpha_{1,2}=0$~\cite{kiss2017mathematics}. 
This behavior is consistent with previous results on higher-order contagion~\cite{burgio2024triadic,malizia2025disentangling}.
To determine the \emph{type} of transition, we perform a center-manifold reduction~\cite{castillo2004dynamical,kuznetsov1998elements}. 
Near the epidemic threshold, the dynamics reduces to the normal form
\begin{equation}
\label{eq:center_manifold_reduction}
\dot{u}=h\,u^2 + z\,\phi\,u + \mathcal{O}(u^3,\phi u^2),
\end{equation}
where $\phi=\lambda_1-\lambda_1^*$ measures the distance from criticality.
The coefficients $h$ and $z$ are given by standard projections of the nonlinear vector field onto the critical eigenspace of the Jacobian (see Appendix), and quantify, respectively, the leading nonlinear self-interaction and the linear unfolding of the instability.
We find that $z>0$ at criticality; therefore, the direction of the bifurcation is entirely controlled by the sign of $h$: $h<0$ yields a supercritical (continuous) transition, while $h>0$ implies a subcritical (backward) bifurcation with bistability and explosive onset.
Although the nonlinear coefficient $h$ can be obtained analytically as a rational function implicitly depending on the critical point $\lambda_1^*$ and the model parameters, its dependence on the key control parameters is not immediately obvious. Consequently, the bistability threshold $\hat{\lambda}_2$ is determined numerically. 
Nevertheless, inspecting the full expansion of the nonlinear coefficient shows that $h$ depends nonlinearly on both the overlap $\alpha_{1,2}$ and the critical pairwise infectivity $\lambda_1^*$, with contributions up to fourth order in $\alpha_{1,2}$.
These terms encode competing reinforcing and suppressing effects arising from pairwise and group interactions.

 \begin{figure}
     \centering
     \includegraphics[width=1\linewidth]{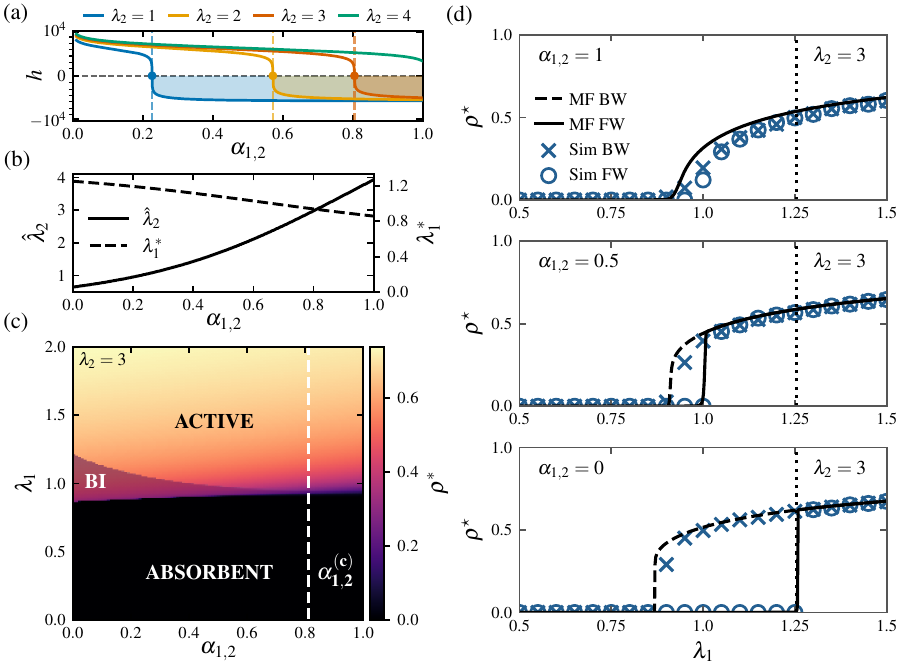}
\caption{\textbf{ Nestedness shapes critical transitions.}
(a) Nonlinear coefficient $h$ from center-manifold theory as a function of $\alpha_{1,2}$, for $\lambda_2\in\{1,2,3,4\}$ (with $k_1=5$, $k_2=2$). Vertical dashed lines mark the critical overlap $\alpha_{1,2}^{(c)}$ at which $h=0$ and the transition changes from subcritical (bistable) to supercritical (continuous).
(b) Critical three-body infectivity $\hat{\lambda}_2$ (solid) required for bistability and the corresponding epidemic threshold $\lambda_1^*$ (dashed), both obtained numerically as functions of $\alpha_{1,2}$ for $k_1=5$, $k_2=2$.
(c) Phase diagram in the $(\alpha_{1,2},\lambda_1)$ plane predicted by the model for $k_1=5$, $k_2=2$, and $\lambda_2=3$, showing that increasing $\alpha_{1,2}$ lowers $\lambda_1^*$ while shrinking the bistable region, which disappears for $\alpha_{1,2}>\alpha_{1,2}^{(c)}$, yielding continuous transitions.
(d) Stationary infected density $\rho^*$ from theory (lines) and Gillespie simulations (markers) on random regular hypergraphs with $N=3000$, $k_1=5$, and $k_2=2$, for three representative values of $\alpha_{1,2}$ at $\lambda_2=3$. For $\alpha_{1,2}=1$, forward (FW) and backward (BW) branches coincide (continuous transition); for $\alpha_{1,2}=0.5$, bistability emerges 
for $\alpha_{1,2}=0$, bistability is maximal and the forward threshold occurs at $\lambda_1^{*,(0)}$ (vertical dotted lines).
}
     \label{fig:fig2}
 \end{figure}

Figure~\ref{fig:fig2}(a) shows that $h$ decreases monotonically with $\alpha_{1,2}$ for $k_1=5$, $k_2=2$ and $\lambda_2\in\{1,2,3,4\}$, identifying a critical overlap $\alpha_{1,2}^{(c)}$ where $h=0$ (dashed): increasing nestedness weakens the nonlinear amplification needed to sustain bistability and eventually turns a backward bifurcation into a continuous onset. For $\lambda_2=4$, $h$ remains positive over $\alpha_{1,2}\in[0,1]$, indicating that bistability survives even at strong overlap, although it is progressively weakened. Together with Eq.~\eqref{eq:transcritical_approximated}, this already reveals the dual role of overlap: it lowers $\lambda_1^*$ while pushing the system away from the subcritical regime.

For $\alpha_{1,2}=0$, the condition for bistability simplifies to:
\begin{equation}
\label{eq:backward_condition_alpha0}
\hat{\lambda}_2^{(0)} = \frac{(k_1 - 1)^2}{k_1^2}.
\end{equation}
Thus, when dyadic and triadic interactions are uncorrelated, the onset of the backward bifurcation is controlled solely by the pairwise connectivity $k_1$. Notably, sparse pairwise layers favor explosive behavior at smaller $\lambda_2$.

Figure~\ref{fig:fig2}(b) reports $\hat{\lambda}_2(\alpha_{1,2})$ (from $h=0$) together with the corresponding $\lambda_1^*(\alpha_{1,2})$ (from the Jacobian), showing that $\hat{\lambda}_2$ increases while $\lambda_1^*$ decreases with overlap. Figure~\ref{fig:fig2}(c) summarizes the stationary states of infected densities $\rho^* \equiv \rho^{\rm I}$ in the ($\lambda_1$,$\alpha_{1,2}$) space, at $\lambda_2=3$: the bistable region shrinks with $\alpha_{1,2}$ and vanishes at $\alpha_{1,2}^{(c)}$. To validate these predictions, we perform Gillespie simulations on random regular hypergraphs with tunable $\alpha_{1,2}$ (SM). Figure~\ref{fig:fig2}(d) shows excellent agreement between theory and simulations. Small deviations at large $\alpha_{1,2}$ are consistent with nestedness-induced pairwise clustering in sparse hypergraphs, which is neglected by our theory~\cite{miller2009spread,miller2009percolation}.

\textit{Microscopic mechanisms underlying the anticipated onset and suppressed bistability.---}
Having established the macroscopic effects of nestedness, we now exploit the analytical tractability of our SIS framework to uncover their microscopic origin. Specifically, we use early-time dynamical correlations to characterize how nestedness reshapes contagion pathways already near the disease-free state.
From Eq.~\eqref{eq:evolution_infecteds}, the infected population grows whenever
\begin{equation}
    \left( \lambda_1 + \lambda_2 \frac{\rho^{\rm ISI_\Delta}}{\rho^{\rm SI}} \right)
    \frac{\rho^{\rm SI}}{\rho^{\rm I}}>1.
\end{equation}
The ratios $\Pi=\rho^{\rm SI}/\rho^{\rm I}$ and $\delta=\rho^{\rm ISI_\Delta}/\rho^{\rm SI}$ act as \emph{fast variables}~\cite{barnard2019epidemic,keeling1999effects}: they relax much faster than $\rho^{\rm I}$ and rapidly reach quasi-stationary values, denoted by $\bar{\Pi}$ and $\bar{\delta}$ (SM). 

Near the disease-free state, the epidemic threshold is entirely determined by these early-time correlations,
\begin{equation}
\label{eq:lambda1_critical_fastvariables}
\lambda_1^*=\frac{1}{\bar{\Pi}}-\lambda_2\bar{\delta}.
\end{equation}
The quasi-stationary fast variables can be obtained analytically as implicit functions of $\lambda_1^*$ (SM). In particular, $\bar{\delta}=0$ when $\alpha_{1,2}=0$, showing that higher-order contagion does not contribute to the epidemic onset in the absence of nestedness. Substituting the analytical fast-variable solutions into Eq.~\eqref{eq:lambda1_critical_fastvariables} exactly recovers the epidemic threshold obtained from the linear stability analysis of the full system, revealing that the onset of contagion is completely encoded in the early-time dynamical correlations generated by nestedness.

To further uncover the microscopic origin of this effect, we decompose the infected density into pairwise- and group-mediated contributions,
\begin{equation}
    \begin{array}{ll}
    \dot{\rho}^{\rm I}_1 = -\mu \rho^{\rm I}_1 + \beta_1 k_1 \rho^{\rm SI}, \\[5pt]
    \dot{\rho}^{\rm I}_2 = -\mu \rho^{\rm I}_2 + \beta_2 k_2 \rho^{\rm ISI_\Delta},
    \end{array}
\end{equation}
where $\rho^{\rm I}_1$ and $\rho^{\rm I}_2$ denote the pairwise- and three-body-mediated contributions to the infected population, respectively. The same decomposition can be applied to all motif variables. Applying the chain rule to the corresponding ratios then yields a closed early-time description in terms of disentangled fast variables, which naturally split into pairwise and higher-order contributions, $\Pi=\Pi_1+\Pi_2$ and $\delta=\delta_1+\delta_2$.

For group-state variables, the pairwise contribution can be further decomposed into external and group-embedded transmission events. For instance, 
$\rho^{\rm ISI_\Delta}_{1}
=
\rho^{\rm ISI_\Delta}_{1,\rm ext}
+
\rho^{\rm ISI_\Delta}_{1,\rm int}$
which leads to the corresponding evolution equations (Appendix). Consequently, the pairwise fast variable can also be decomposed as $\delta_1=\delta_{1,\rm ext}+\delta_{1,\rm int}$,
allowing us to quantify separately the contributions of external and embedded transmission routes at the early-stage (Appendix).

\begin{figure}[t!]
     \centering
     \includegraphics[width=1\linewidth]{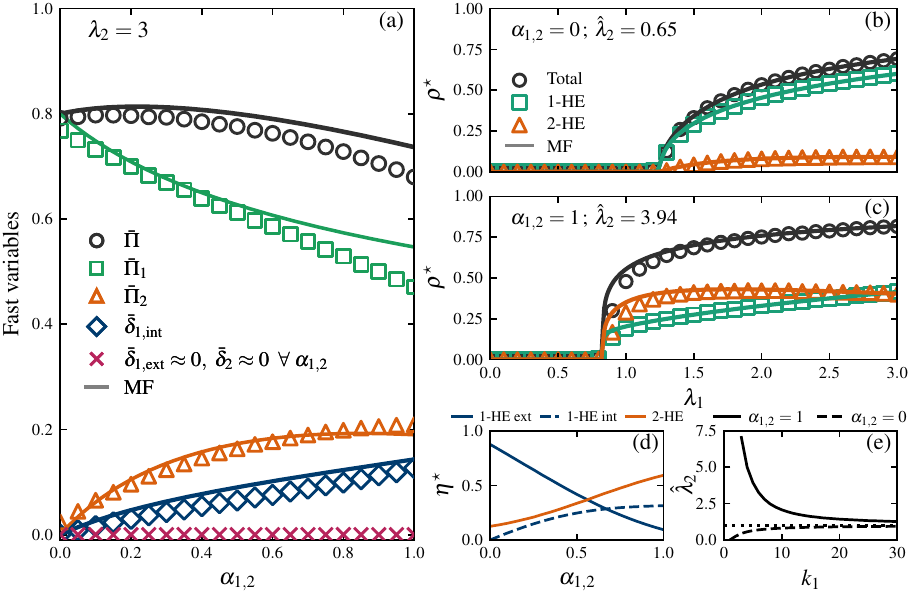}
\caption{\textbf{Microscopic mechanisms underlying the anticipated onset and suppressed bistability.} (a) Quasi-stationary states for the fast variables $\Pi$ and $\delta$ as functions of $\alpha_{1,2}$, obtained from theory (lines) and Gillespie simulations (symbols) on random regular hypergraphs with $N=3000$, $k_1=5$, $k_2=2$, and $\lambda_2=3$. Each point is evaluated close to $\lambda_1^*$ for each $\alpha_{1,2}$. The contribution to $\bar{\delta}$ originates entirely from the internal pairwise component $\delta_{1,\rm int}$ and increases with $\alpha_{1,2}$.
(b,c) Decomposition of the stationary infected density $\rho^*$ into total, pairwise (1-hyperedge), and group-based (2-hyperedge) contributions for $\alpha_{1,2}=0$ and $\alpha_{1,2}=1$ on the same hypergraphs, evaluated at the corresponding critical values $\hat{\lambda}_2$, before bistability emerges. In both cases the transition remains continuous, although nested structures display a sharper onset.
(d) Fractions of accumulated infection events $\eta_1^*$ (pairwise, internal and external) and $\eta_2^*$ (higher-order) near $\lambda_1^*$ as functions of $\alpha_{1,2}$, showing a progressive shift from external pairwise to internal and higher-order transmission channels.
(e) Critical group infectivity $\hat{\lambda}_2$ versus $k_1$ (with $k_2=2$) for $\alpha_{1,2}=0$ (dashed, analytical) and $\alpha_{1,2}=1$ (solid, numerical), illustrating the strong suppression of explosive behavior induced by overlap in sparse networks; both curves converge to the mean-field limit $\hat{\lambda}_2=1$ (dotted line) for large $k_1$.}
     \label{fig:fig3}
 \end{figure}

Figure~\ref{fig:fig3}(a) compares the theoretical predictions obtained from this fast-variable framework with Gillespie simulations. Fast variables are measured by initializing the system from a single infected node near $\lambda_1^*$ and averaging over $5000$ realizations within the early-time window $0.005<\rho^{\rm I}(t)<0.01$. Theory and simulations show excellent agreement, particularly for the higher-order contribution $\bar{\Pi}_2$ and the internal pairwise contribution $\bar{\delta}_{1,\rm int}$.
Most importantly, the disentangled dynamics reveals that the only nonzero contribution to $\bar{\delta}$ originates from \emph{internal} pairwise transmission within higher-order groups, and increases monotonically with nestedness. Combined with Eq.~\eqref{eq:lambda1_critical_fastvariables}, this demonstrates that nestedness anticipates the epidemic threshold by promoting mixed group configurations already at the earliest stages of the dynamics, thereby increasing the ratio $\rho^{\rm ISI_\Delta}/\rho^{\rm SI}$. 

We now connect this early-time picture to the stationary regime by decomposing $\rho^*$ into total, pairwise, and group-based contributions. Figures~\ref{fig:fig3}(b,c) show this decomposition for $\alpha_{1,2}=0$ and $\alpha_{1,2}=1$ at their respective $\hat{\lambda}_2$ values. While pairwise transmission dominates for $\alpha_{1,2}=0$, increasing nestedness enhances internal pairwise transmission, activating more higher-order interactions and shifting the dominant transmission channel without inducing bistability.
To quantify this route redistribution, we define $\eta_1^*$ and $\eta_2^*$ as the fractions of accumulated pairwise and higher-order infection events up to stationarity, with $\eta_1^*=\eta_{1,\rm ext}^*+\eta_{1,\rm int}^*$. Figure~\ref{fig:fig3}(d) completes the microscopic picture by showing these quantities, averaged over $1000$ Gillespie realizations on random regular hypergraphs with $N=3000$, $k_1=5$, and $k_2=2$, evaluated at $\lambda_1=\lambda_1^*+0.05$ and the corresponding bistability threshold $\hat{\lambda}_2$.
As nestedness increases, infection events are progressively redirected from external pairwise links to internal pairwise and higher-order transmission. This redistribution enhances higher-order activity, but at the same time concentrates pairwise and higher-order contagion within the same groups, making the two transmission channels increasingly redundant. Consistently, center-manifold theory predicts an increasing $\hat{\lambda}_2(\alpha_{1,2})$, showing that progressively stronger higher-order transmission is required to compensate for this loss of independent nonlinear reinforcement and sustain bistability.

Figure~\ref{fig:fig3}(e) illustrates how nestedness fundamentally changes the role of pairwise connectivity. For $\alpha_{1,2}=0$, the analytical predictions of Eqs.~\eqref{eq:transcritical_approximated} and \eqref{eq:backward_condition_alpha0} explain the dashed curve: low $k_1$ delays the epidemic onset while lowering the critical value $\hat{\lambda}_2$ required for bistability, consistently with the general picture of discontinuous transitions driven by nonlinear feedback~\cite{thiele2019first}.  Nestedness reverses this dependence. Equation~\eqref{eq:transcritical_approximated} predicts that its correction to $\lambda_1^*$ is strongest for small $k_1$. Together with the fast-variable result that group-embedded pairwise routes are the only contribution anticipating activation, this explains why the solid curve is shifted upward at low $k_1$: sparse systems contain fewer external routes, making nestedness most effective at anticipating activation while simultaneously suppressing the independent nonlinear feedback. As $k_1$ increases, external routes are progressively restored and $\hat{\lambda}_2$ decreases, with both curves approaching $\hat{\lambda}_2\to1$ for large $k_1$~\cite{iacopini2019simplicial}. Together, these results show that the onset and the nature of the transition are complementary signatures of the same microscopic route redistribution.

\begin{figure}[t!]
    \centering
    \includegraphics[width=1\linewidth]{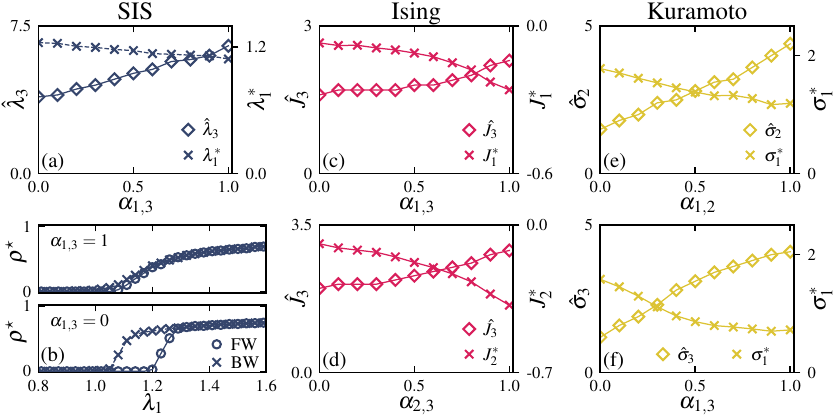}
\caption{\textbf{Generality across interaction orders and dynamical processes.}
Critical onset thresholds and higher-order parameters required for a backward bifurcation as functions of nestedness, extracted from the forward and backward branches of the corresponding stationary order parameters. Results are shown for (a) SIS dynamics with pairwise and four-body interactions; (b) representative forward (FW) and backward (BW) SIS branches for $\alpha_{1,3}=0$ and $\alpha_{1,3}=1$ at $\lambda_3=6$; (c) Ising dynamics with pairwise and four-body interactions; (d) Ising dynamics with three- and four-body interactions only; and Kuramoto dynamics with (e) pairwise--three-body and (f) pairwise--four-body interactions. In all cases, nestedness shifts the onset of collective behavior to lower values of the corresponding control parameter while increasing the critical higher-order parameter required for discontinuous behavior.}

    \label{fig:fig4}
    \vspace{-0.2cm}
\end{figure}

\textit{A general mechanism across interaction orders and dynamical processes.---}
Having established the mechanism for contagion with pairwise and three-body groups, we now investigate its generality across interaction orders and dynamical processes. In direct analogy with Fig.~\ref{fig:fig2}(b), for each system we compute the corresponding stationary order parameter and use its forward and backward branches to estimate the lower-order coupling marking collective onset and the higher-order coupling at which a robust hysteresis loop first emerges, according to a common operational criterion (SM). We first consider SIS dynamics on random regular hypergraphs with pairwise and four-body interactions ($N=2000$, $k_1=7$, $k_3=2$), with tunable nestedness $\alpha_{1,3}$ (SM). Susceptible nodes are infected through pairwise interactions at rate $\beta_1$ or through four-body interactions at rate $\beta_3$ when the other three nodes are infected, with $\lambda_3=k_3\beta_3/\mu$. Figure~\ref{fig:fig4}(a) shows that increasing $\alpha_{1,3}$ lowers the epidemic threshold $\lambda_1^*$ while increasing the critical four-body infectivity $\hat{\lambda}_3$ required for a backward bifurcation, demonstrating that the dual effect persists beyond three-body contagion.

We next consider the higher-order Ising model of Ref.~\cite{robiglio2025higher}, where pairwise, three-body, and four-body interactions favor aligned spin configurations through couplings $J_1$, $J_2$, and $J_3$. The system undergoes a transition from a paramagnetic (disordered) to a ferromagnetic (ordered) phase, quantified by the stationary absolute magnetization $|\langle m\rangle|$ (Appendix). Since discontinuous transitions arise only for $m>2$, this model also allows us to probe nestedness entirely within the higher-order sector. We therefore consider (i) pairwise and four-body interactions, estimating $J_1^*$ and $\hat{J}_3$, and (ii) three- and four-body interactions, extracting $J_2^*$ and $\hat{J}_3$. For hypergraphs with pairwise and four-body interactions ($N=2000$, $k_1=7$, $k_3=2$), Figure~\ref{fig:fig4}(c) shows that increasing $\alpha_{1,3}$ lowers $J_1^*$ while increasing the critical four-body coupling $\hat{J}_3$ required for hysteresis. Furthermore, for hypergraphs with three- and four-body interactions only, with an empty set of dyadic interactions ($N=2000$, $k_2=7$, $k_3=2$) and tunable nestedness $\alpha_{2,3}$ (SM), increasing $\alpha_{2,3}$ again lowers the transition point while increasing $\hat{J}_3$ [Fig.~\ref{fig:fig4}(d)]. The same mechanism therefore persists even without pairwise interactions.

Finally, we consider the higher-order Kuramoto model of Ref.~\cite{skardal2020higher}, where oscillators interact through order-dependent couplings $\sigma_m$. Synchronization is quantified by the stationary time-averaged Kuramoto order parameter $\langle r\rangle$ (Appendix). We consider (i) pairwise and three-body interactions, extracting $\sigma_1^*$ and $\hat{\sigma}_2$, and (ii) pairwise and four-body interactions, estimating $\sigma_1^*$ and $\hat{\sigma}_3$. In both cases, increasing nestedness lowers the synchronization threshold while shifting the onset of bistability toward larger higher-order couplings [Fig.~\ref{fig:fig4}(e,f)]. The SM provides an extensive analysis across interaction orders and dynamical processes, and further shows that the Kuramoto results are robust to alternative higher-order coupling functions. Overall, these results show that nestedness promotes collective onset while suppressing discontinuous transitions across fundamentally different dynamical systems.

\textit{Conclusions.---}
In this paper, we showed that nestedness between interaction orders plays a dual role in collective dynamics: it promotes the onset of collective behavior while suppressing the nonlinear feedback responsible for abrupt transitions. Using higher-order contagion as an analytically tractable setting, we developed a microscopic theory revealing that nestedness redistributes dynamical routes from external neighborhoods to group-embedded ones, reducing the structural independence between interaction orders.
This route redistribution lowers activation thresholds but progressively weakens the independent reinforcement required for bistability, thereby smoothing discontinuous transitions. Beyond contagion dynamics, we demonstrated that the same mechanism persists across multiple interaction orders and dynamical processes, including higher-order Ising model and Kuramoto synchronization, and even in the absence of pairwise interactions.
Taken together, our results identify nestedness as a general structural mechanism governing both the onset and the nature of collective transitions in higher-order systems.

\textit{Appendix A: Homogeneous mean-field model.---} To analytically characterize the impact of nestedness on SIS dynamics, we develop a homogeneous mean-field description for regular hypergraphs with pairwise (1-hyperedges) and three-body (2-hyperedges) interactions. Susceptible nodes become infected through pairwise interactions at rate $\beta_1$ or through a three-body interaction containing two infected neighbors at rate $\beta_2$, while infected nodes recover at rate $\mu$. Throughout this Appendix, we denote the inter-order overlap by $\alpha\equiv\alpha_{1,2}$.

The model explicitly retains the cross-order dynamical correlations induced by nestedness by tracking the state vector $\mathbf{x}
=
(\rho^{\rm I},
\rho^{\rm SI},
\rho^{\rm SSS_\Delta},
\rho^{\rm SSI_\Delta},
\rho^{\rm ISI_\Delta})$,
where
$\rho^{\rm A}=[A]/N$,
$\rho^{\rm AB}=[AB]/(Nk_1)$,
and
$\rho^{\rm ABC_\Delta}=[ABC_\Delta]/(2Nk_2)$
denote normalized node, link, and three-body motif densities, respectively. Homogeneity implies
$\rho^{\rm SI}=\rho^{\rm IS}$ and
$\rho^{\rm ISI_\Delta}
=
\rho^{\rm IIS_\Delta}
=
\rho^{\rm SII_\Delta}$.
The evolution equations naturally generate higher-order composite motifs coupling links and three-body interactions. Their explicit definitions are reported in the Supplemental Material (SM).
The resulting dynamical system reads
\begin{equation}
\label{eq:model_system_eqs}
    \begin{array}{ll}
         \displaystyle \dot{\rho}^{\rm I} &= -\mu \rho^{\rm I} + \beta_1 k_1 \rho^{\rm SI} + \beta_2 k_2 \rho^{\rm ISI_\Delta}; \\ [10pt]

\dot{\rho}^{\rm SI} &= \beta_1 \left[ \left( k_1 - 1\right) \left(\rho^{\rm SSI} - \rho^{\rm ISI}\right) - \rho^{\rm SI}  \right]  \\[5pt] &  + \beta_2\frac{k_2}{k_1} \left[\left(k_1 - 2\alpha\right) \left(\rho^{\rm IIS_\Delta S} -\rho^{\rm IIS_\Delta I} \right) -2 \alpha \rho^{\rm ISI_\Delta} \right] \\[5pt] & + \mu \left(\rho^{\rm II} - \rho^{\rm SI}\right); \\ [10pt]

\dot{\rho}^{\rm SSS_\Delta} &= -3\beta_1\left(k_1-2\alpha \right) \rho^{\rm SSS_\Delta I} - 3\beta_2\left(k_2-1 \right)\rho^{\rm SSSII_{\bowtie}} \\[5pt] & + 3\mu\rho^{\rm SSI_\Delta}; \\ [10pt]

    \dot{\rho}^{\rm SSI_\Delta} &= \beta_1 \left[\left(k_1 -2\alpha\right) \left( \rho^{\rm SSS_\Delta I} -2 \rho^{\rm ISS_\Delta I}\right) - 2\alpha \rho^{\rm SSI_\Delta}\right] \\[5pt] & + \beta_2 \left(k_2 -1\right) \left( \rho^{\rm SSSII_{_{\bowtie}}} - 2\rho^{\rm ISSII_{\bowtie}} \right) \\[5pt] &+ \mu \left( 2\rho^{\rm ISI_\Delta} - \rho^{\rm SSI_\Delta}\right); 
    \\ [10pt]
    
         \dot{\rho}^{\rm ISI_\Delta} &= \beta_1 \left[\left(k_1-2\alpha\right) \left(2 \rho^{\rm ISS_\Delta I } - \rho^{\rm IIS_\Delta I} \right)\right] \\[5pt]& + 2\alpha \beta_1 \left( \rho^{\rm SSI_\Delta} - \rho^{\rm ISI_\Delta}\right) \\[5pt] & +\beta_2 \left[\left( k_2 - 1\right) \left( 2 \rho^{\rm ISSII_{\bowtie}} - \rho^{\rm IISII_{\bowtie }}\right) - \rho^{\rm ISI_\Delta} \right] \\ [5pt] &
          + \mu \left(\rho^{\rm III_\Delta} - 2 \rho^{\rm ISI_\Delta}\right)
         , 
    \end{array}
\end{equation}

The remaining motif densities are obtained from the conservation identities
$\rho^{\rm S}=1-\rho^{\rm I}$,
$\rho^{\rm II}=\rho^{\rm I}-\rho^{\rm SI}$,
$\rho^{\rm SS}=\rho^{\rm S}-\rho^{\rm SI}$,
and
$\rho^{\rm III_\Delta}
=
1-\rho^{\rm SSS_\Delta}
-
3\rho^{\rm SSI_\Delta}
-
3\rho^{\rm ISI_\Delta}$.

Nestedness enters the model solely through the availability of external pairwise transmission routes. For a node already belonging to a three-body interaction, increasing nestedness progressively replaces external links by group-embedded ones, yielding an effective number of external links $k_{\rm 1,ext}= (1-\alpha_{1,2})k_1 + \alpha_{1,2}(k_1 -2) = k_1-2\alpha_{1,2}$, which appears whenever transmission requires an external infected neighbor while the focal susceptible node is already conditioned on belonging to a three-body interaction.

The hierarchy is closed through standard homogeneous approximations~\cite{kiss2017mathematics,house2009motif}, neglecting pairwise clustering and intra-order overlap among distinct three-body interactions~\cite{malizia2025hyperedge}. Consequently, higher-order motifs are expressed in terms of lower-order densities while explicitly retaining the cross-order correlations generated by nestedness. The complete set of closure relations and their derivation are reported in the SM.

To determine the onset of spreading, we introduce the rescaled infectivities $\lambda_1=k_1\beta_1/\mu$ and $\lambda_2=k_2\beta_2/\mu$ and linearize Eqs.~\eqref{eq:model_system_eqs} around the disease-free equilibrium $\mathbf{x}^*=(0,0,1,0,0)$. The epidemic threshold (transcritical bifurcation) is obtained by imposing that the dominant eigenvalue of the Jacobian crosses zero. For compactness, we define $\Gamma(\lambda_1)\equiv k_1-(k_1-1)\lambda_1$ and $\Theta(\alpha,\lambda_1)\equiv k_1k_2+(k_1+k_2-1)\lambda_1- k_1 - 4\alpha k_2\lambda_1$, which yields the exact condition
\begin{equation}
\label{eq:epidemic_threshold_exact}
\alpha \lambda_1\lambda_2\Big[4\alpha^2 k_2\lambda_1+k_1\,\Theta(\alpha,\lambda_1)\Big]
= k_1^2k_2\,\Gamma(\lambda_1).
\end{equation}

\smallskip
\noindent\textit{Appendix B: Center-manifold reduction.—}
To characterize the \emph{type} of transition at the epidemic threshold, we perform a center-manifold reduction of the closed system $\dot{\mathbf{x}}=\mathbf{F}(\mathbf{x};\lambda_1,\lambda_2,\alpha)$ about the disease-free equilibrium $\mathbf{x}^*=(0,0,1,0,0)$. The critical point is defined implicitly by Eq.~\eqref{eq:epidemic_threshold_exact}; throughout, all quantities below are evaluated on this critical manifold. Let $J=D_{\mathbf{x}}\mathbf{F}(\mathbf{x}^*;\lambda_1,\lambda_2,\alpha)\big|_{\rm crit}$ be the Jacobian at criticality and let $\mathbf{w}$ and $\mathbf{v}$ denote the right and left eigenvectors associated with the simple zero eigenvalue, normalized by $\mathbf{v}^\top\mathbf{w}=1$. Introducing the unfolding parameter $\phi=\lambda_1-\lambda_1^*$ (with $\lambda_1^*$ defined implicitly by Eq.~\eqref{eq:epidemic_threshold_exact}), the dynamics on the center manifold reduces to the scalar normal form $\dot{u}=h u^2 + z\,\phi\,u + \mathcal{O}(u^3,\phi u^2)$.

In terms of the vector field $\mathbf{F}$, the quadratic coefficient is
\begin{equation}
\label{eq:a_coefficient}
h=\sum_{i,j,k=1}^{b} v_i\,
\frac{\partial^2 F_i}{\partial x_j\,\partial x_k}\!\left(\mathbf{x}^*;\lambda_1,\lambda_2,\alpha\right)\Big|_{\phi=0}\,
w_j w_k,
\end{equation}
and the parameter-dependent coefficient is
\begin{equation}
\label{eq:b_coefficient}
z=\sum_{i,j=1}^{b} v_i\,
\frac{\partial^2 F_i}{\partial x_j\,\partial \lambda_1}\!\left(\mathbf{x}^*;\lambda_1,\lambda_2,\alpha\right)\Big|_{\phi=0}\,
w_j,
\end{equation}
with $b=\dim(\mathbf{x})=5$.

\noindent\textit{Appendix C: Fast-variable system.—}
In the early stage ($\rho^{\rm I}\to 0$), ratios of motif densities relax on a fast time scale. We define the fast variables
$\Pi\equiv\rho^{\rm SI}/\rho^{\rm I}$ and $\Psi\equiv\rho^{\rm ISI_\Delta}/\rho^{\rm I}$, so that $\delta\equiv\rho^{\rm ISI_\Delta}/\rho^{\rm SI}=\Psi/\Pi$ (main text).
To close the fast-variable dynamics we introduce two additional ratios, $\Omega\equiv\rho^{\rm SSS_\Delta}/\rho^{\rm I}$ and
$\Upsilon\equiv\rho^{\rm SSI_\Delta}/\rho^{\rm I}$.
The remaining pair fast variable is eliminated using the identity
$\rho^{\rm SI}+\rho^{\rm II}=\rho^{\rm I}$, i.e., $\rho^{\rm II}/\rho^{\rm I}=1-\Pi$.
Fast-variable equations follow by differentiating ratios via the chain rule; e.g.,
$\dot{\Pi}=(\dot{\rho}^{\rm SI}/\rho^{\rm I})-\Pi(\dot{\rho}^{\rm I}/\rho^{\rm I})$, and similarly for $\Upsilon,\Psi,\Omega$. 

Substituting the motif equations and closures into the chain-rule expressions yields

\begin{equation}
\label{eq:fastvars_system}
\begin{aligned}
\dot{\Pi} =& \mu(1-\Pi) + \beta_1\left(k_1-2\right) \Pi
          + \beta_2\frac{k_2}{k_1}(k_1-4\alpha)\Psi \\
          &- \beta_1 k_1 \Pi^2 - \beta_2 k_2 \Pi\Psi,\\
\dot{\Upsilon} =& \beta_1(k_1-2\alpha)\Pi - 2\beta_1\alpha\,\Upsilon
             + \big(2\mu+\beta_2(k_2-1)\big)\Psi \\&
             - \beta_1 k_1 \Pi\Upsilon - \beta_2 k_2 \Upsilon\Psi,\\
\dot{\Omega} =& \mu (2\Upsilon+\Omega)-3\beta_1(k_1 -               2\alpha)\Pi - 3\beta_2(k_2-1)\Psi 
            \\ &- \beta_1  k_1 \Pi \Omega -\beta_2 k_2 \Omega \Psi . \\
\dot{\Psi} =& 2\beta_1\alpha\,\Upsilon +                            \big(-2\beta_1\alpha-\mu-\beta_2\big)\Psi +             \mu\,\big(1-\Omega\\&
            -3\Upsilon-3\Psi\big)- \beta_1 k_1 \Pi\Psi - \beta_2 k_2  \Psi^2,\\
\end{aligned}
\end{equation}

This system directly provides the early-time evolution of $\Psi=\rho^{\rm ISI_\Delta}/\rho^{\rm I}$ and hence of
$\delta=\Psi/\Pi=\rho^{\rm ISI_\Delta}/\rho^{\rm SI}$.

\textit{Appendix D: Disentangling by interaction order and internal/external pairwise channels.---}
We decompose the evolution equations for the state densities in Eq.~\eqref{eq:model_system_eqs} by explicitly disentangling the the microscopic processes contributing to each transition. In particular, we distinguish between (i) infections generated by pairwise interactions (subscript $1$) and (ii) infections mediated by three-body interactions (subscript $2$). For group-state densities, such as $\rho^{\rm ISI_\Delta}$, the pairwise contribution can be further separated into transmission along interactions embedded within three-body groups and transmission along external pairwise interactions (subscripts $1,\mathrm{int}$ and $1,\mathrm{ext}$, respectively), yielding an expanded system of differential equations. For example, for the state variable $\rho^{\rm ISI_\Delta}$, we obtain:

\begin{equation}
    \begin{array}{ll}
         \dot{\rho}^{\rm ISI_\Delta}_{1,\rm ext} =& + \mu \left( \rho^{\rm III_\Delta}_{1,\rm ext} - 2 \rho^{\rm ISI_\Delta}_{1,\rm ext}\right) \\[5pt] &+ \beta_1 \left( k_1 - 2\alpha \right) \left ( 2\rho^{\rm ISS_\Delta I }- \rho^{\rm IIS_\Delta I }\right);  \\ [5pt]
        \dot{\rho}^{\rm ISI_\Delta}_{1,\rm int} =& + \mu \left( \rho^{\rm III_\Delta}_{1,\rm int} - 2 \rho^{\rm ISI_\Delta}_{1,\rm int}\right) + 2 \alpha \beta_1 \left( \rho^{\rm SSI_\Delta} - 2 \rho^{\rm ISI_\Delta}\right); \\[5pt]
        \dot{\rho}^{\rm ISI_\Delta}_{2} =&  + \mu \left( \rho^{\rm III_\Delta}_{2} - 2 \rho^{\rm ISI_\Delta}_{2}\right) \\[5pt]& +\beta_2 \left[\left( k_2 - 1\right) \left( 2 \rho^{\rm ISSII_{\bowtie}} - \rho^{\rm IISII_{\bowtie }}\right) - \rho^{\rm ISI_\Delta} \right]
    \end{array}
    \label{eq:groups_disentangled_int_ext}
\end{equation}

The full disentangled system of equations can be found in the SM. Furthermore, the fast variables can be consistently decomposed into the sum of their higher-order, internal pairwise, and external pairwise contributions as
\begin{equation}
    \begin{array}{l}
    \Upsilon=\Upsilon_2+\Upsilon_{1,{\rm int}}+\Upsilon_{1,{\rm ext}};\quad
    \Psi=\Psi_2+\Psi_{1,{\rm int}}+\Psi_{1,{\rm ext}}; \\[5pt]
    \Omega=\Omega_2+\Omega_{1,{\rm int}}+\Omega_{1,{\rm ext}}.
    \end{array}
\end{equation}
Their corresponding differential equations are also displayed in the SM. This disentangled system provides a closed early-time description of the fast variables, capturing the contributions by interaction order and by internal/external link channel within groups.

\textit{Appendix E: Higher-order Ising model.---}
To demonstrate the generality of our results beyond spreading dynamics, we consider the higher-order Ising model introduced in Refs.~\cite{robiglio2025higher,son2026phase}. Each node carries a binary spin $s_i\in\{-1,+1\}$, and an energetic contribution is assigned whenever all spins belonging to the same hyperedge are aligned. The Hamiltonian reads
\begin{equation}
H=
-\sum_{\ell=1}^{\ell_{\max}}
J_\ell
\sum_{\{\sigma:|\sigma|=\ell+1\}}
\delta\!\left(\{s_i\}_{i\in\sigma}\right),
\label{eq:hamiltonian_ising}
\end{equation}
where $J_\ell$ is the coupling associated with $(\ell+1)$-body interactions, and
\begin{equation}
\delta(s_1,\ldots,s_n)=
\begin{cases}
1 & \text{if } s_1=s_2=\cdots=s_n,\\
0 & \text{otherwise},
\end{cases}
\end{equation}
is the generalized Kronecker delta. 
We characterize the state of the system via the absolute value of the stationary magnetization,
\begin{equation}
  |\langle m\rangle|,
\qquad
m=\frac{1}{N}\sum_{i=1}^{N}s_i,  
\end{equation}
where $\langle\cdot\rangle$ denotes the temporal average over Monte Carlo samples collected in the stationary regime. Details on the stochastic simulations are provided in the SM.

\textit{Appendix F: Kuramoto dynamics.---}
We consider Kuramoto oscillator dynamics with group interactions up to order $M=3$ \cite{skardal2020higher}, where $\theta_i$ is the phase of oscillator $i$, $\omega_i$ its natural frequency, $\sigma_m$ the coupling strength of interactions of order $m$, and $k_m$ the corresponding regular degree. The adjacency tensors $a^{(1)}_{ij}$, $a^{(2)}_{ijl}$, and $a^{(3)}_{ijln}$ encode pairwise, three-body, and four-body interactions, respectively. The dynamics is given by
\begin{equation}
    \begin{array}{ll}
    \dot{\theta}_i =& \omega_i + \dfrac{\sigma_1}{k_1} \sum\limits_{j=1}^N a_{ij}^{(1)} \sin \left( \theta_j - \theta_i \right) \\
    &+ \dfrac{\sigma_2}{k_2} \sum\limits_{j=1}^N  \sum\limits_{l=1}^N  a_{ijl}^{(2)} \sin \left( 2\theta_j - \theta_l - \theta_i \right) \\
    & + \dfrac{\sigma_3}{k_3} \sum\limits_{j=1}^N  \sum\limits_{l=1}^N \sum\limits_{n=1}^N  a_{ijln}^{(3)} \sin \left( \theta_j + \theta_l - \theta_n - \theta_i \right).
    \end{array}
\end{equation}
Synchronization is quantified by the Kuramoto order parameter
$r(t)=\left|N^{-1}\sum_{j=1}^N e^{i\theta_j(t)}\right|$, ranging from $0$ (incoherence) to $1$ (phase synchronization). In the main text, we vary the nestedness between interaction orders and identify the critical pairwise coupling $\sigma_1^*$ for the onset of synchronization and the critical higher-order coupling $\hat{\sigma}_m$ ($m=2,3$) required for explosive synchronization from the stationary time average $\langle r\rangle$ of $r(t)$ (SM).

\smallskip
\textit{Acknowledgments.---}F.M. acknowledges support from the Austrian Science Fund (FWF) through project 10.55776/PAT1652425. A.G. acknowledges the PhD studentship support from Northeastern University London. F.B. acknowledges support from the Austrian Science Fund (FWF) through project 10.55776/PAT1052824 and project 10.55776/PAT1652425.

\bibliographystyle{unsrt}
\bibliography{biblio}

\clearpage
\onecolumngrid

\renewcommand{\figurename}{Supplementary Figure}
\renewcommand{\tablename}{Supplementary Table}

\setcounter{secnumdepth}{2}

\setcounter{section}{0}
\renewcommand{\thesection}{S\arabic{section}}

\setcounter{subsection}{0}
\renewcommand{\thesubsection}{\thesection.\arabic{subsection}}

\setcounter{equation}{0}
\renewcommand{\theequation}{S\arabic{equation}}

\begin{center}
  \textbf{\large Supplemental Material for: \\ \vspace{0.25cm}
Nested hyperedges promote the onset of collective transitions but suppress explosive behavior \vspace{0.25cm}} \\[.2cm]

  Federico Malizia,$^1$ \,Andr\'es Guzm\'an,$^2$ Federico Battiston,$^{1,3,*}$ and Istv\'an Z. Kiss$^{2,4,*}$ \\ [.1cm]
  {\itshape ${}^1$ Department of Network and Data Science, Central European University, Vienna, Austria\\
            ${}^2$ Network Science Institute, Northeastern University London, London E1W 1LP, United Kingdom\\
            ${}^3$ Department of AI, Data and Decision Sciences, Luiss University of Rome, Viale Romania, Rome, Italy\\
            ${}^4$Department of Mathematics, Northeastern University, Boston, MA 02115, USA \\
            }
\end{center}

\makeatletter
\def\@thefnmark{*}
\@footnotetext{These authors jointly supervised this work.}
\makeatother

\onecolumngrid

\section{Generality across interaction orders and dynamical processes}
In the main text, we analytically characterize the effect of nestedness for SIS dynamics with pairwise and three-body interactions. Figure~4 of the main text shows that the same qualitative phenomenology extends beyond this minimal setting: nestedness anticipates the onset of collective behavior while suppressing bistability or hysteresis across larger interaction groups and different dynamical processes.

In this section, we provide additional details on the analyses presented in Fig.~4 of the main text. We first describe the common numerical procedure used to estimate the critical quantities from simulations and then discuss, separately, the SIS extension to larger interaction groups, the higher-order Ising model, and Kuramoto dynamics.

\subsection{Numerical estimation of critical values}
In the main text, we analytically characterize the effect of nestedness for SIS dynamics with pairwise and three-body interactions. As shown in Fig.~4 of the main text, however, the same qualitative phenomenology extends to systems with larger interaction groups and to fundamentally different dynamical processes, for which no analogous analytical framework is currently available. In this section, we describe the common numerical procedure used to estimate the critical quantities reported in Fig.~4.

The purpose of this procedure is not to determine the exact bifurcation points of each individual model, but rather to provide a robust and uniform benchmark allowing a quantitative comparison of the effect of nestedness across different systems. In all cases, the extracted quantities characterize transitions that are already fully developed and are therefore only weakly affected by finite-size fluctuations, stochastic variability, and small numerical oscillations close to the bifurcation.

For every dynamical process, every value of the nestedness parameter, and every value of the higher-order coupling, we compute forward and backward continuations of the stationary order parameter. The forward branch is obtained by gradually increasing the lower-order control parameter, whereas the backward branch is obtained by decreasing it from an ordered, active, or synchronized initial condition. The simulation protocol specific to each dynamical process is described in the following subsections.

Whenever multiple realizations are performed, the stationary order parameter is first averaged over the independent realizations associated with the same point in parameter space. In the SIS model, this average is conditioned on surviving realizations in order to remove stochastic extinctions. We denote the resulting ensemble-averaged stationary forward and backward branches by $O_{\rm FW}(y_m,y_n,\alpha)$ and $O_{\rm BW}(y_m,y_n,\alpha)$, respectively, where $y_m$ and $y_n$ are control parameters for each order of interaction, with $m<n$.

To identify the onset of hysteresis, we first compute, for every value of the higher-order coupling,
\[
\Delta O(y_m;y_n)
=
\max\!\left\{
O_{\rm BW}(y_m;y_n)-O_{\rm FW}(y_m;y_n),\,0
\right\},
\]
and define the corresponding hysteresis-loop area as
\begin{equation}
A_{\rm H}(y_n)
=
\int
\Delta O(y_m;y_n)\,
{\rm d}x,
\label{eq:hysteresis_area}
\end{equation}
which is evaluated numerically by trapezoidal integration over the sampled values of the lower-order control parameter.

The critical higher-order coupling $\hat y_n$ is then identified as the smallest sampled value satisfying
\[
A_{\rm H}(y_n)>A_{\rm thr},
\]
provided that a genuine hysteresis loop is present. Specifically, we additionally require the forward branch to remain below a prescribed threshold while the backward branch simultaneously remains above the same threshold for at least three consecutive values of the lower-order control parameter. This condition prevents spurious detections caused by finite-size fluctuations or stochastic noise.

Once $\hat y_n$ has been determined, the corresponding onset of collective behavior, denoted by $y_m^*$, is estimated from the forward branch as the first value of the lower-order control parameter for which the stationary order parameter exceeds a prescribed reference level. The crossing is evaluated by linear interpolation between consecutive sampled points.

Finally, to quantify the overall extent of the bistable phase, we compute the area of the bistable region in the two-dimensional phase diagram
$A_{\rm BI}$,
obtained by integrating the width of the hysteresis region over the sampled values of the higher-order control parameter. Throughout the manuscript, we report the normalized quantity $A_{\rm BI}/A_{\rm tot}$, where $A_{\rm tot}$ denotes the total explored parameter-space area.

To ensure consistency across all dynamical processes, we adopt the same operational criteria throughout. For SIS, Ising, and Kuramoto dynamics, we use threshold values $\rho_{\rm thr}^*=0.01$, $\langle m\rangle_{\rm thr}=0.15$, and $\langle r\rangle_{\rm thr}=0.10$, respectively. The same threshold is employed to define the genuine separation between the forward and backward branches. The hysteresis-loop threshold is fixed to $A_{\rm thr}=0.02$ for all dynamical processes, and the genuine-gap condition always requires at least three consecutive sampled points satisfying the threshold criterion.

For SIS dynamics, realizations ending in stochastic extinction (stationary prevalence smaller than $10^{-3}$) are discarded before averaging; if no realization survives, the stationary prevalence is set to zero. For the Ising and Kuramoto models, the stationary order parameter is averaged over all independent realizations. In the Kuramoto simulations, forward and backward continuations are sampled on different grids of the lower-order coupling; therefore, both branches are linearly interpolated onto their common parameter grid before evaluating Eq.~(\ref{eq:hysteresis_area}), without extrapolation outside the simulated range.

This procedure provides the numerical estimates of $\lambda_1^*$ and $\hat{\lambda}_3$ for SIS dynamics, $J_1^*$, $J_2^*$, and $\hat J_3$ for the Ising model, and $\sigma_1^*$, $\hat\sigma_2$, and $\hat\sigma_3$ for Kuramoto dynamics reported in Fig.~4 of the main text.
\subsection{SIS dynamics beyond pairwise and three-body interactions}
In this section, we provide additional details on the effect of nestedness in SIS dynamics beyond the minimal setting considered in the main text, where the analytical framework focuses on systems with pairwise and three-body interactions ($M=2$). We investigate whether the same phenomenology persists for larger interaction groups by considering two complementary scenarios.

We first study systems composed of pairwise and four-body interactions only, corresponding to hyperedges of orders $m=1$ and $m=3$. In this case, the set of three-body interactions ($m=2$) is empty, allowing us to isolate the effect of the nestedness between pairwise and four-body interactions, $\alpha_{1,3}$.

We then consider a second setting including three-body and four-body interactions, where pairwise interactions are also present. This example illustrates that nestedness between three-body and four-body interactions, quantified by $\alpha_{2,3}$, also influences the collective behavior of the epidemic dynamics. The SIS dynamics follows the same microscopic rules introduced in the main text. In systems with interactions up to order $M=3$, a susceptible node can become infected through pairwise interactions at rate $\beta_1$, through a three-body interaction at rate $\beta_2$ when the other two nodes in the hyperedge are infected, or through a four-body interaction at rate $\beta_3$ when the other three nodes in the hyperedge are infected. Infected nodes recover independently at rate $\mu$, which we set to $\mu=1$ throughout this work. As in the main text, we introduce the rescaled infection rates $\lambda_m=k_m\beta_m/\mu$.

For the first scenario, we consider regular hypergraphs composed only of pairwise and four-body interactions, with $N=2000$, $k_1=7$, and $k_3=2$, while the set of three-body interactions is empty. We generate ensembles with $11$ values of the nestedness parameter $\alpha_{1,3}$, uniformly spanning the interval from the non-nested configuration ($\alpha_{1,3}=0$) to the fully nested one ($\alpha_{1,3}=1$). The details of the hypergraph-generation procedure are provided in Sec.~\ref{sec:rewiring_procedure}.

\begin{figure}[h!]
    \centering
    \includegraphics[width=\linewidth]{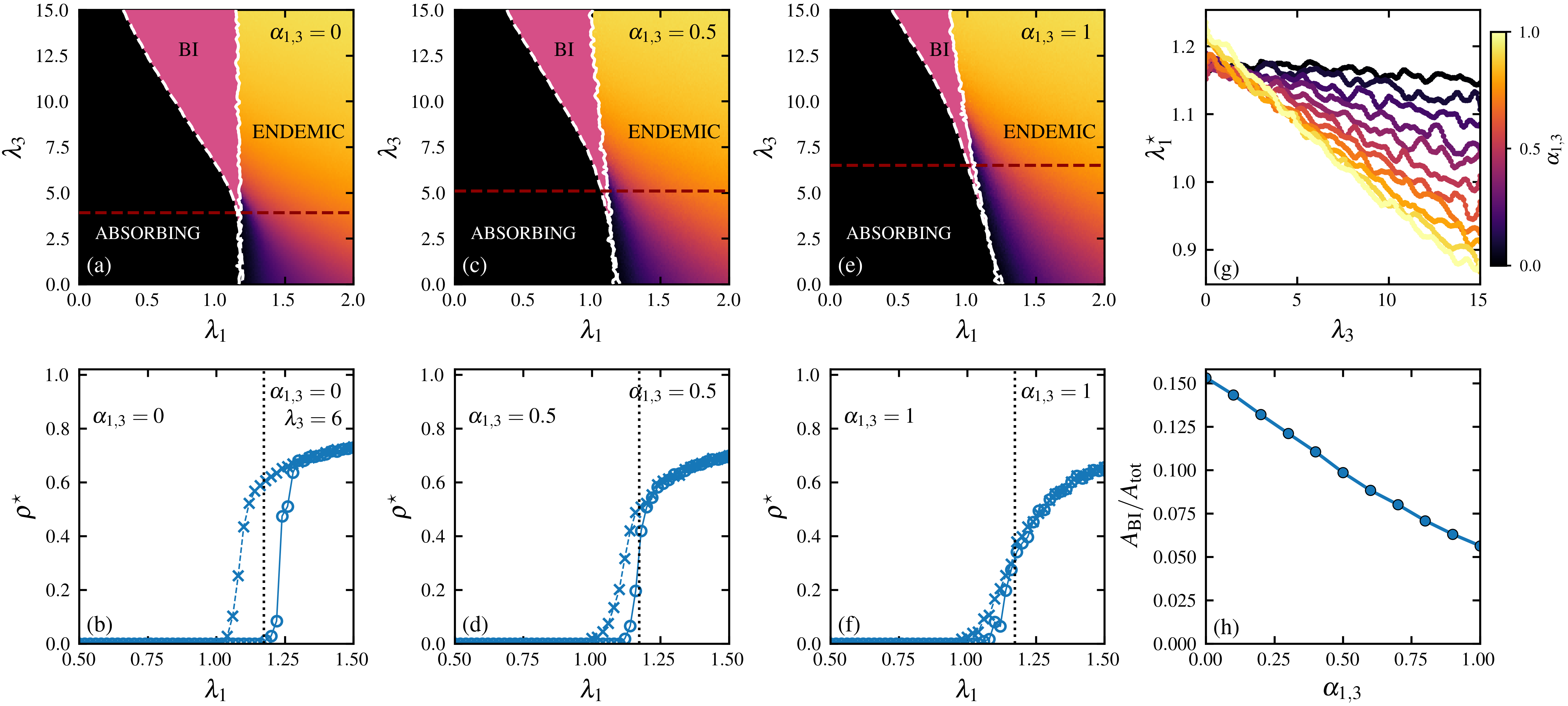}
\caption{
\textbf{Effect of nestedness on SIS dynamics with pairwise and four-body interactions.}
Panels (a,c,e) show the phase diagrams in the $(\lambda_1,\lambda_3)$ plane for three representative values of the nestedness parameter $\alpha_{1,3}$. The white solid line denotes the epidemic onset obtained from the forward branch, separating the absorbing and active phases, while the white dashed line marks the backward transition delimiting the bistable region (BI). The red dashed horizontal lines indicate the critical higher-order infectivity $\hat{\lambda}_3$, numerically extracted as reported in Fig.~4 of the main text. Panels (b,d,f) report the corresponding forward (circles) and backward (crosses) stationary prevalence $\rho^*$ as functions of $\lambda_1$, illustrating the progressive anticipation of the epidemic onset and the suppression of hysteresis as nestedness increases. The vertical dotted lines denote the numerically extracted critical value $\lambda_1^*$ for $\alpha_{1,3} = 0$. Panel (g) shows the dependence of $\lambda_1^*$ on the higher-order infectivity $\lambda_3$ for different values of $\alpha_{1,3}$, while panel (h) reports the fraction of the phase diagram occupied by the bistable region, $A_{\mathrm{BI}}/A_{\mathrm{tot}}$, as a function of nestedness. Increasing nestedness systematically shifts the epidemic onset towards lower values of $\lambda_1$ while reducing the extent of the bistable region, confirming that the microscopic mechanism identified analytically in the main text persists beyond pairwise and three-body interactions.} \label{fig:SIS_13_sm}
\end{figure}

The results are reported in Supplementary Figure~\ref{fig:SIS_13_sm}. For each value of the nestedness parameter $\alpha_{1,3}$, we numerically reconstructed the phase diagram in the $(\lambda_1,\lambda_3)$ plane by performing forward and backward continuations over a grid of coupling values. Specifically, we considered $11$ equally spaced values of $\alpha_{1,3}$ between $0$ and $1$. For every point of the phase diagram, we simulated $100$ independent forward realizations, initialized with a fraction $10^{-3}$ of infected nodes, and $50$ backward realizations, initialized from a highly endemic configuration with an initial prevalence $\rho(0)=0.9$. The stationary prevalence was then estimated by averaging over all surviving realizations, as described in the previous section.

The results are shown in Fig.~\ref{fig:SIS_13_sm}. Panels (a,c,e) report the phase diagrams in the $(\lambda_1,\lambda_3)$ plane for three representative values of the nestedness parameter, $\alpha_{1,3}=0$, $0.5$, and $1$. The solid white curve denotes the epidemic onset obtained from the forward continuation, separating the absorbing and endemic phases, while the dashed white curve identifies the backward transition delimiting the bistable region, where both phases coexist. The red horizontal dashed lines indicate the numerically estimated values of the critical higher-order infectivity $\hat{\lambda}_3$, obtained according to the operational procedure described in the previous subsection.

The phase diagrams clearly reveal the dual effect of nestedness. In the absence of nestedness [$\alpha_{1,3}=0$, panel (a)], the bistable region occupies a broad portion of the parameter space, indicating that explosive epidemic transitions occur over a wide range of four-body infectivities. Moreover, consistently with the analytical prediction obtained for pairwise and three-body interactions [Eq.~(2) of the main text], the epidemic threshold is nearly independent of the higher-order infectivity, apart from the small finite-size and stochastic fluctuations visible close to the transition. As nestedness increases, the bistable region progressively shrinks, while the epidemic threshold becomes increasingly sensitive to the four-body infectivity, demonstrating that nestedness substantially amplifies the effect of higher-order interactions on the onset of epidemic spreading.

The same behavior is illustrated from a complementary perspective in panels (b,d,f), which report the forward and backward stationary prevalence as functions of $\lambda_1$ for a fixed higher-order infectivity $\lambda_3=6$. While a pronounced hysteresis loop is observed for $\alpha_{1,3}=0$, increasing nestedness progressively reduces the separation between the two branches, as highlighted for $\alpha_{1,3}=1$.

Unlike the analytical threshold $\hat{\lambda}_2$ derived in Fig.~2 of the main text, the quantity $\hat{\lambda}_3$ does not represent the exact point at which the transition changes from continuous to discontinuous. Since no analytical theory is currently available for this system, $\hat{\lambda}_3$ should instead be interpreted as a numerical benchmark marking the emergence of a robust hysteresis loop. Although approximate, this definition provides a consistent criterion for comparing different values of nestedness and quantifying how the onset of bistability evolves beyond the analytically tractable setting considered in the main text.

Panels (g) and (h) summarize these trends over the entire parameter space. Panel (g) reports the epidemic threshold $\lambda_1^*$ as a function of the higher-order infectivity $\lambda_3$ for the $11$ values of the nestedness parameter $\alpha_{1,3}$. Consistently with the analytical results obtained for pairwise and three-body interactions, increasing nestedness systematically anticipates the epidemic onset, shifting the critical pairwise infectivity towards lower values. Although finite-size and stochastic fluctuations introduce small irregularities in the numerically extracted curves, the dependence of $\lambda_1^*$ on $\lambda_3$ becomes progressively stronger as nestedness increases.

Finally, panel (h) quantifies the suppression of hysteresis by reporting the normalized bistable area $A_{\rm BI}/A_{\rm tot}$, where $A_{\rm BI}$ is the area enclosed between the forward and backward transition lines and $A_{\rm tot}$ is the total explored parameter-space area. Both quantities are evaluated numerically from the transition lines shown in panels (a,c,e). The relative bistable area decreases monotonically with increasing nestedness, demonstrating that the suppression of bistability predicted analytically for pairwise and three-body interactions extends to systems with four-body interactions.

We now consider the second setting of our analysis, aimed at investigating whether nestedness between higher-order interactions alone can also affect the collective dynamics. Having established that nestedness between pairwise and higher-order interactions strongly anticipates the epidemic onset and suppresses bistability, we next examine the role of nestedness between three-body and four-body interactions.

To this end, we consider random regular hypergraphs with $N=2000$, $k_1=12$, $k_2=7$, and $k_3=2$. We generate ensembles with tunable nestedness $\alpha_{2,3}$ between three-body and four-body interactions, spanning the full range from the non-nested configuration ($\alpha_{2,3}=0$) to the fully nested one ($\alpha_{2,3}=1$). Again, the details of the procedure to minimize $\alpha_{2,3}$ are provided in Sec.~S\ref{sec:rewiring_procedure}.
Pairwise interactions are then generated independently as an uncorrelated random regular network, yielding negligible overlap with both higher-order interaction sets ($\alpha_{1,2}\simeq0$ and $\alpha_{1,3}\simeq0$). This construction isolates the effect of nestedness between three-body and four-body interactions while suppressing the influence of pairwise-higher-order correlations.

\begin{figure}
    \centering
    \includegraphics[width=1\linewidth]{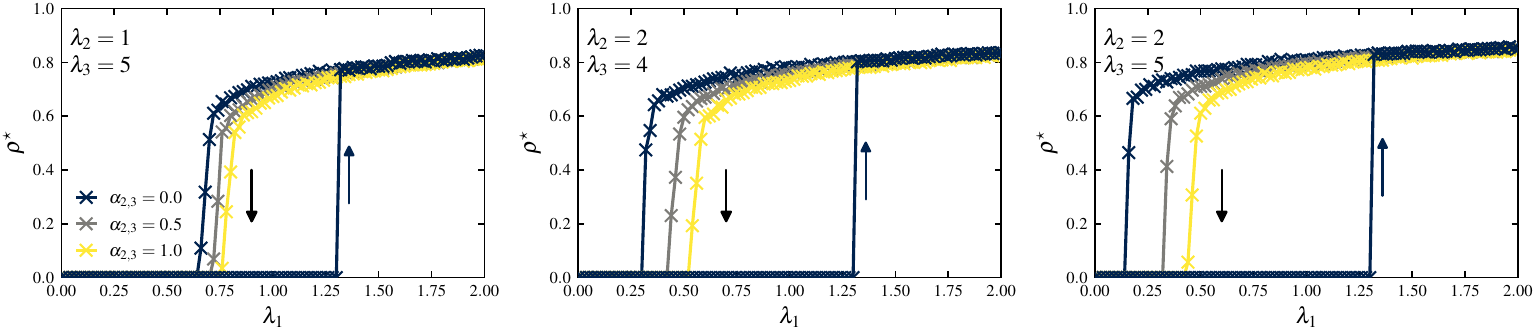}
    \caption{\textbf{Effect of nestedness between three-body and four-body interactions on SIS dynamics.}
Stationary prevalence $\rho^\star$ as a function of the pairwise infectivity $\lambda_1$ for three representative values of the nestedness parameter $\alpha_{2,3}$, at fixed higher-order infectivities $\lambda_2$ and $\lambda_3$. Forward (circles) and backward (crosses) continuations are obtained by increasing and decreasing $\lambda_1$, respectively. Since pairwise interactions are generated independently, with $\alpha_{1,2}\simeq0$ and $\alpha_{1,3}\simeq0$, the forward epidemic threshold remains approximately unchanged as $\alpha_{2,3}$ varies. By contrast, increasing nestedness between three-body and four-body interactions shifts the backward transition toward larger values of $\lambda_1$, thereby reducing the width of the hysteresis loop and the associated bistable region.}
    \label{fig:SM_sis_123}
\end{figure}

The results are shown in Supplementary Figure \ref{fig:SM_sis_123}. Since pairwise interactions are generated independently of the higher-order structure, with $\alpha_{1,2}\simeq0$ and $\alpha_{1,3}\simeq0$, changing $\alpha_{2,3}$ produces essentially no variation in the forward epidemic threshold. This is consistent with the mechanism identified in the main text, where the anticipation of the onset originates from the embedding of lower-order interactions within larger groups.
Nevertheless, increasing the nestedness between three-body and four-body interactions still affects the discontinuous transition. While the forward branch remains almost unchanged, the backward branch progressively shifts towards larger values of $\lambda_1$, reducing the width of the hysteresis loop. As a consequence, the bistable region becomes progressively smaller as $\alpha_{2,3}$ increases.

Although considerably weaker than the effect induced by nestedness between pairwise and higher-order interactions, these results show that correlations entirely within the higher-order interaction structure also contribute to shaping the collective transition. In particular, they further suppress bistability, suggesting that nestedness between groups of different sizes acts as a general mechanism reducing the structural independence between interaction channels, even when pairwise interactions do not directly participate in the nested organization.

In the following sections of the Supplemental Material, we provide a detailed analysis of two additional classes of dynamical processes, namely the higher-order Ising model and higher-order Kuramoto dynamics, further illustrating the generality of the phenomenology reported in Fig.~4 of the main text.

\subsection{Higher-order Ising dynamics}
To investigate whether the effect of nestedness extends beyond spreading dynamics, we next consider the higher-order Ising model introduced in Refs.~\cite{robiglio2025higher,son2026phase}. Unlike the standard $p$-spin Hamiltonian, where the interaction energy depends on the product of the spins within a group, this model assigns an energetic contribution only when all spins belonging to the same hyperedge are aligned. This construction preserves the $\mathbb{Z}_2$ symmetry of the classical Ising model while providing a natural extension to arbitrary interaction orders~\cite{robiglio2025higher}. The Hamiltonian is
\begin{equation}
H=
-\sum_{\ell=1}^{\ell_{\max}}
J_\ell
\sum_{\{\sigma:|\sigma|=\ell+1\}}
\delta\!\left(\{s_i\}_{i\in\sigma}\right),
\label{eq:hamiltonian_ising}
\end{equation}
where $s_i\in\{-1,+1\}$ denotes the spin associated with node $i$, $J_\ell$ is the coupling strength associated with $(\ell+1)$-body interactions, and
\begin{equation}
\delta(s_1,\ldots,s_n)=
\begin{cases}
1, & s_1=\cdots=s_n,\\
0, & \text{otherwise},
\end{cases}
\end{equation}
is the generalized Kronecker delta. Consequently, each hyperedge contributes to the energy only when all spins belonging to the same interaction group share the same orientation, independently of whether they are all in the $+1$ or $-1$ state.

The equilibrium dynamics is sampled using a standard Metropolis--Hastings Monte Carlo algorithm. At each Monte Carlo step, we perform $N$ randomly selected spin, and the flip is proposed and accepted with probability
\begin{equation}
P_{\rm acc}
=
\min\!\left\{
1,
e^{-\beta\Delta H}
\right\},
\end{equation}
where $\Delta H$ is the corresponding energy variation and $\beta=1/T$ is the inverse temperature. Throughout this work we fix $\beta=1$ and use the coupling strengths as the control parameters.

As order parameter we consider the absolute value of the stationary magnetization,
\begin{equation}
    |\langle m\rangle|,
\qquad
m=\frac{1}{N}\sum_{i=1}^{N}s_i,
\end{equation}
where $\langle\cdot\rangle$ denotes the average over stationary Monte Carlo samples collected after the equilibration period. The absolute value accounts for the two symmetry-related ferromagnetic equilibrium states.

The system undergoes a transition between a disordered paramagnetic phase and an ordered ferromagnetic phase. A distinctive feature of the Hamiltonian in Eq.~\eqref{eq:hamiltonian_ising} is that discontinuous transitions, accompanied by bistability and hysteresis, emerge only when interactions of order $m>2$ are present, namely from four-body interactions onward.

This property makes the model particularly suitable for assessing the generality of the mechanism identified for SIS dynamics. Unlike epidemic spreading, where higher-order interactions alone can already induce discontinuous transitions, the Hamiltonian in Eq.~\eqref{eq:hamiltonian_ising} allows us to disentangle the role of nestedness between pairwise and four-body interactions from that of nestedness entirely within the higher-order sector. Accordingly, we consider the  two distinct structural settings.

For case (I), we study regular hypergraphs composed of pairwise and four-body interactions only, with $N=2000$, $k_1=7$, and $k_3=2$, while the set of three-body interactions is empty. Hypergraphs are generated with $11$ uniformly spaced values of the nestedness parameter $\alpha_{1,3}\in[0,1]$. For each value of $\alpha_{1,3}$, we reconstruct the phase diagram in the $(J_1,J_3)$ plane by computing forward and backward continuations for $21$ values of the four-body coupling $J_3$. Forward simulations are initialized from random spin configurations, whereas backward simulations start from the fully ordered state with $s_i=+1$ for every node. Each branch is averaged over $50$ independent realizations, from which the critical values $J_1^*$ and $\hat{J}_3$ are extracted following the numerical procedure described in the previous section.

\begin{figure}[t!]
    \centering
    \includegraphics[width=\linewidth]{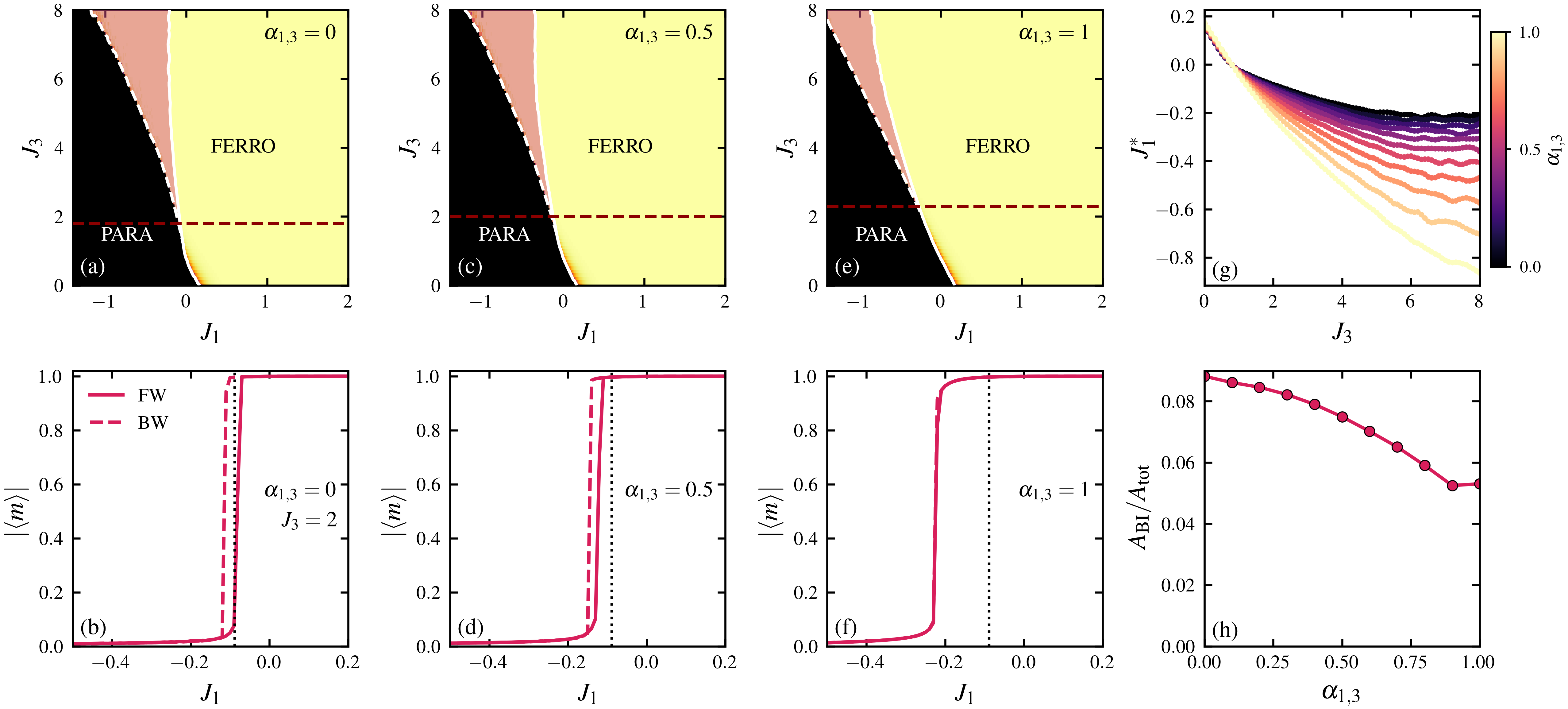}
\caption{
\textbf{Effect of nestedness on higher-order Ising dynamics with pairwise and four-body interactions.}
Panels (a,c,e) show the phase diagrams in the $(J_1,J_3)$ plane for three representative values of the nestedness parameter $\alpha_{1,3}$. The white solid line denotes the forward transition separating the paramagnetic (PARA) and ferromagnetic (FERRO) phases, while the white dashed line marks the backward transition delimiting the bistable region (BI). The red dashed horizontal lines indicate the critical four-body coupling $\hat{J}_3$, numerically extracted following the procedure adopted in Fig.~4 of the main text. Panels (b,d,f) report the corresponding forward (solid) and backward (dashed) stationary magnetization $|\langle m\rangle|$ as functions of $J_1$, illustrating the progressive anticipation of the ferromagnetic transition and the suppression of hysteresis as nestedness increases. The vertical dotted lines denote the numerically extracted critical value $J_1^*$ for $\alpha_{1,3} = 0$. Panel (g) shows the dependence of $J_1^*$ on the four-body coupling $J_3$ for different values of $\alpha_{1,3}$, while panel (h) reports the fraction of the phase diagram occupied by the bistable region, $A_{\mathrm{BI}}/A_{\mathrm{tot}}$, as a function of nestedness. Increasing nestedness systematically shifts the ferromagnetic transition towards lower values of $J_1$ while reducing the extent of the bistable region, demonstrating that the phenomenology identified analytically for SIS dynamics also extends to equilibrium higher-order spin systems.}

    \label{fig:sm_ising_13}
\end{figure}

The results for case (I) are reported in Supplementary Fig.~\ref{fig:sm_ising_13}. Panels (a,c,e) show the phase diagrams in the $(J_1,J_3)$ plane for three representative values of the nestedness parameter, $\alpha_{1,3}=0$, $0.5$, and $1$. The solid and dashed white curves identify the forward and backward transition lines, respectively, separating the paramagnetic and ferromagnetic phases and delimiting the bistable region. The red dashed horizontal lines indicate the numerically estimated values of $\hat{J}_3$, namely the minimum four-body coupling required for the emergence of a robust hysteresis loop.

An important difference with respect to the SIS model is already visible in the non-nested case. Consistently with previous results for the higher-order Ising model~\cite{robiglio2025higher,son2026phase}, the paramagnetic--ferromagnetic transition depends strongly on the four-body coupling even when the two interaction orders are structurally uncorrelated, $\alpha_{1,3}=0$. Thus, higher-order interactions shift the transition independently of nestedness. Nestedness, however, substantially amplifies this dependence. As shown in panel (g), the critical pairwise coupling $J_1^*$ decreases with $J_3$ for every value of $\alpha_{1,3}$, but this dependence becomes progressively stronger as nestedness increases. In other words, the same four-body coupling produces a substantially larger shift of the transition when pairwise interactions are preferentially embedded within four-body groups.

Panels (b,d,f) provide representative forward and backward continuations at fixed $J_3=2$ for the same three values of $\alpha_{1,3}$. The separation between the two branches progressively decreases as nestedness increases, showing directly how nestedness suppresses the hysteretic regime. This behavior is confirmed over the entire phase diagram in panel (h), where the normalized bistable area $A_{\rm BI}/A_{\rm tot}$ decreases systematically with $\alpha_{1,3}$.

\begin{figure}[h!]
    \centering
    \includegraphics[width=\linewidth]{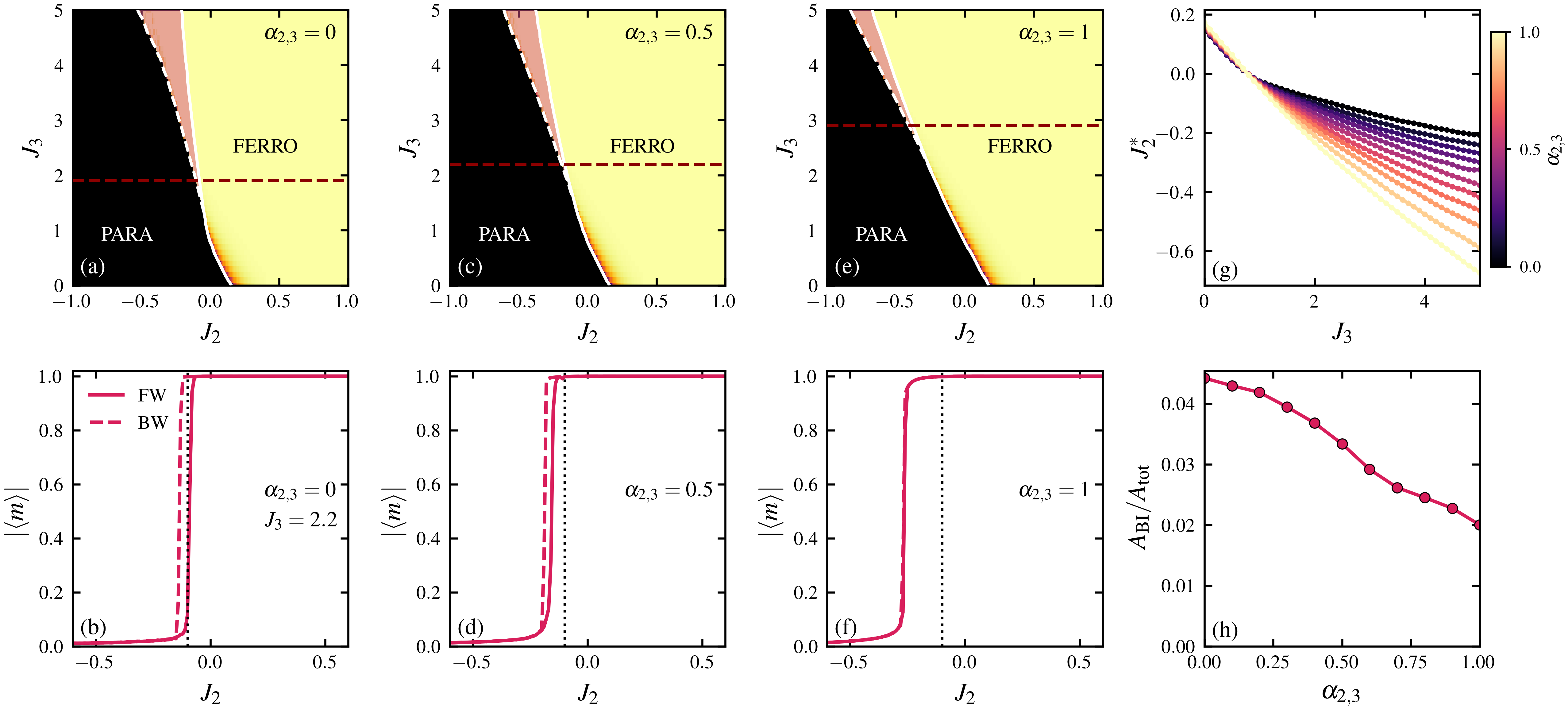}
\caption{
\textbf{Effect of nestedness on higher-order Ising dynamics with three-body and four-body interactions.}
Panels (a,c,e) show the phase diagrams in the $(J_2,J_3)$ plane for three representative values of the nestedness parameter $\alpha_{2,3}$. The white solid line denotes the forward transition separating the paramagnetic (PARA) and ferromagnetic (FERRO) phases, while the white dashed line marks the backward transition delimiting the bistable region (BI). The red dashed horizontal lines indicate the critical four-body coupling $\hat{J}_3$, numerically extracted following the procedure adopted in Fig.~4 of the main text. Panels (b,d,f) report the corresponding forward (solid) and backward (dashed) stationary magnetization $|\langle m\rangle|$ as functions of $J_2$. The vertical dotted lines denote the numerically extracted critical value $J_2^*$ for $\alpha_{2,3}=0$. Panel (g) shows the dependence of $J_2^*$ on the four-body coupling $J_3$ for different values of $\alpha_{2,3}$, while panel (h) reports the fraction of the phase diagram occupied by the bistable region, $A_{\mathrm{BI}}/A_{\mathrm{tot}}$, as a function of nestedness. As in the case of pairwise and four-body interactions, increasing nestedness anticipates the onset of the ordered phase while reducing the extent of the bistable region, showing that the same qualitative effect persists for nestedness entirely within the higher-order interaction structure.
}
    \label{fig:sm_ising_23}
\end{figure}

We now turn to case (II), where the system contains three-body and four-body interactions only, and nestedness is controlled through $\alpha_{2,3}$. The corresponding results are shown in Supplementary Figure~\ref{fig:sm_ising_23}. Panels (a,c,e) report the phase diagrams in the $(J_2,J_3)$ plane for $\alpha_{2,3}=0$, $0.5$, and $1$. As before, the solid and dashed white lines denote the forward and backward transition lines, respectively, while the red dashed horizontal lines indicate the numerically estimated critical four-body coupling $\hat{J}_3$.
The phase diagrams show that nestedness between three-body and four-body interactions produces the same qualitative effect observed in the pairwise--four-body setting. The critical three-body coupling $J_2^*$ depends on $J_3$ even in the non-nested case, but this dependence becomes progressively stronger as $\alpha_{2,3}$ increases. This is quantified in panel (g), where $J_2^*$ is reported as a function of $J_3$ for all values of nestedness. Increasing $\alpha_{2,3}$ shifts the transition towards lower values of $J_2$, showing that embedding three-body interactions within four-body groups enhances the contribution of the higher-order coupling to the emergence of collective order.
Again, the same qualitative picture emerges. As shown in panels (b,d,f) for $J_3 = 2.2$, increasing the nestedness between three-body and four-body interactions progressively reduces the separation between the forward and backward branches, indicating a gradual suppression of the hysteretic regime. Accordingly, the numerically estimated critical coupling $\hat{J}_3$ shifts towards larger values, implying that increasingly stronger four-body interactions are required for bistability to develop. This trend is confirmed over the entire parameter space in panel (h), where the normalized bistable area $A_{\rm BI}/A_{\rm tot}$ decreases monotonically with $\alpha_{2,3}$.

Taken together, the two settings reveal that the role of nestedness is not limited to the embedding of pairwise interactions within larger groups. The same phenomenology is observed when nestedness occurs entirely within higher-order interactions, demonstrating that the mechanism is intrinsic to the hierarchical organization of interactions rather than to the specific interaction order involved.
Therefore, although the higher-order Ising model belongs to a fundamentally different class of dynamical processes from epidemic spreading, and although its transition already depends on the higher-order coupling even in the absence of structural correlations, nestedness produces the same dual effect identified for SIS dynamics. Specifically, it enhances the influence of higher-order interactions on the onset of collective ordering while simultaneously suppressing the extent of the bistable region. In the following section, we show that the same phenomenology also emerges in higher-order Kuramoto dynamics, further supporting the generality of this mechanism across qualitatively different collective processes.

\subsection{Higher-order Kuramoto dynamics}

We conclude our analysis by considering synchronization dynamics, providing a third and fundamentally different class of collective processes. Together with the SIS and higher-order Ising models discussed above, this allows us to assess whether the effect of nestedness represents a general feature of collective dynamics on hypergraphs, rather than a property specific to contagion or equilibrium spin systems.

We first consider the higher-order Kuramoto model of Ref.~\cite{skardal2020higher}, which exhibits explosive synchronization and hysteresis driven by higher-order interactions. After characterizing the effect of nestedness within this framework, we repeat the same analysis using an alternative higher-order interaction rule, showing that the phenomenology discussed throughout this work does not depend on the particular microscopic form of the coupling function.

For interactions up to order $M=3$, the dynamics of oscillator $i$ is governed by
\begin{equation}
\begin{aligned}
\dot{\theta}_i
=&\,
\omega_i
+\frac{\sigma_1}{k_1}
\sum_{j=1}^{N}
a_{ij}^{(1)}
\sin(\theta_j-\theta_i)
\\
&
+\frac{\sigma_2}{k_2}
\sum_{j=1}^{N}
\sum_{l=1}^{N}
a_{ijl}^{(2)}
\sin(2\theta_j-\theta_l-\theta_i)
\\
&
+\frac{\sigma_3}{k_3}
\sum_{j=1}^{N}
\sum_{l=1}^{N}
\sum_{n=1}^{N}
a_{ijln}^{(3)}
\sin(\theta_j+\theta_l-\theta_n-\theta_i),
\label{eq:kuramoto_skardal}
\end{aligned}
\end{equation}
where $\theta_i$ and $\omega_i$ denote the phase and natural frequency of oscillator $i$, respectively. The coupling strengths associated with pairwise, three-body, and four-body interactions are denoted by $\sigma_1$, $\sigma_2$, and $\sigma_3$, while $k_1$, $k_2$, and $k_3$ are the corresponding regular degrees. The adjacency tensors $a^{(1)}_{ij}$, $a^{(2)}_{ijl}$, and $a^{(3)}_{ijln}$ encode the pairwise, three-body, and four-body interactions of the underlying hypergraph.

The level of synchronization is quantified through the Kuramoto order parameter
\begin{equation}
r(t)=
\left|
\frac{1}{N}
\sum_{j=1}^{N}
e^{i\theta_j(t)}
\right|,
\label{eq:kuramoto_order_parameter}
\end{equation}
where $r(t) \simeq0$ corresponds to the incoherent state and $r(t)\simeq1$ to complete phase synchronization.

In the case of Kuramoto dynamics, we consider two different interaction settings. In case (I), we study regular random hypergraphs with pairwise and three-body interactions only, with $N=600$, $k_1=5$, and $k_2=2$, while the set of four-body interactions is empty. Hypergraphs are generated with $11$ uniformly spaced values of the nestedness parameter $\alpha_{1,2}\in[0,1]$. In case (II), we instead consider regular random hypergraphs with pairwise and four-body interactions only, with $N=600$, $k_1=7$, and $k_3=2$, while the set of three-body interactions is empty. Again, we generate $11$ values of the nestedness parameter $\alpha_{1,3}\in[0,1]$ using the rewiring procedure described in Sec.~S\ref{sec:rewiring_procedure}.

The equations are integrated using a fourth-order Runge--Kutta scheme. For each value of the nestedness parameter and higher-order coupling, we perform $100$ independent forward continuations and $50$ independent backward continuations. Forward runs are initialized with phases uniformly distributed in $[0,2\pi]$, whereas backward runs start from phases narrowly distributed around the synchronized state. The forward and backward branches are obtained by adiabatically increasing and decreasing the pairwise coupling $\sigma_1$, respectively, using the final state reached at the previous value of $\sigma_1$ as the initial condition for the subsequent simulation. For each realization, we compute the stationary time average $\langle r\rangle$ of $r(t)$ after the transient, and then average $\langle r\rangle$ over the independent realizations of the corresponding forward or backward protocol. From the resulting mean forward and backward branches, we extract the critical values $\sigma_1^*$, $\hat{\sigma}_2$, and $\hat{\sigma}_3$ according to the common numerical procedure described in the previous section.

\begin{figure}
    \centering
    \includegraphics[width=\linewidth]{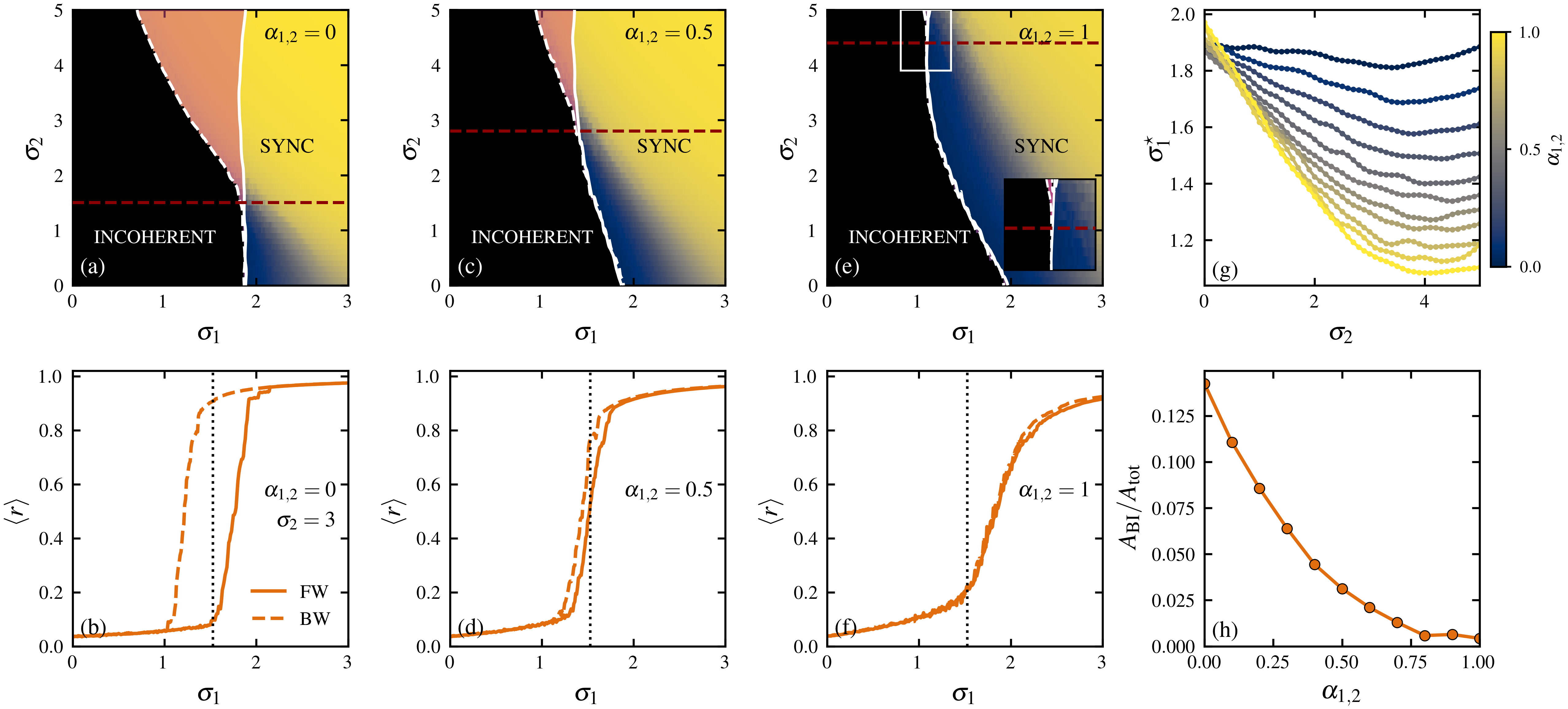}
\caption{
\textbf{Effect of nestedness on Kuramoto dynamics with pairwise and three-body interactions.}
Panels (a,c,e) show the phase diagrams in the $(\sigma_1,\sigma_2)$ plane for three representative values of the nestedness parameter $\alpha_{1,2}$. The white solid line denotes the synchronization onset obtained from the forward branch, separating the incoherent and synchronized phases, while the white dashed line marks the backward transition delimiting the bistable region. The red dashed horizontal lines indicate the critical higher-order coupling $\hat{\sigma}_2$, numerically extracted as reported in Fig.~4 of the main text. The inset in panel (e) highlights the small bistable region emerging immediately above the red dashed line marking the numerically estimated critical higher-order coupling $\hat{\sigma}_2$. Panels (b,d,f) report the corresponding forward (solid) and backward (dashed) stationary synchronization order parameter $r^\star$ as a function of $\sigma_1$, illustrating the anticipation of the synchronization onset and the progressive suppression of hysteresis as nestedness increases. The vertical dotted lines denote the numerically extracted critical value $\sigma_1^*$ for $\alpha_{1,2}=0$. Panel (g) shows the dependence of $\sigma_1^*$ on the higher-order coupling $\sigma_2$ for different values of $\alpha_{1,2}$, while panel (h) reports the fraction of the phase diagram occupied by the bistable region, $A_{\mathrm{BI}}/A_{\mathrm{tot}}$, as a function of nestedness. Increasing nestedness systematically lowers the synchronization threshold while reducing the extent of the bistable region, confirming that the dual effect of nestedness extends to synchronization dynamics with three-body interactions.
}    \label{fig:sm_kuramoto_12}
\end{figure}

We first consider case (I), namely synchronization on random regular hypergraphs with pairwise and three-body interactions only. We generate hypergraphs with $N=600$, $k_1=5$, and $k_2=2$, while the set of four-body interactions is empty. As in the previous sections, we construct $11$ hypergraphs with tunable nestedness $\alpha_{1,2}\in[0,1]$ using the rewiring procedure described in Sec.~S\ref{sec:rewiring_procedure}. For each value of $\alpha_{1,2}$, we compute the synchronization diagrams in the $(\sigma_1,\sigma_2)$ plane by integrating the forward and backward continuations for different values of the higher-order coupling.

The corresponding results are reported in Supplementary Figure~\ref{fig:sm_kuramoto_12}. Panels (a,c,e) show the phase diagrams in the $(\sigma_1,\sigma_2)$ plane for three representative values of the nestedness parameter, $\alpha_{1,2}=0$, $0.5$, and $1$. The solid and dashed white curves denote the forward and backward synchronization transitions, respectively, separating the incoherent and synchronized phases and delimiting the bistable region. The red horizontal dashed lines indicate the numerically estimated values of $\hat{\sigma}_2$, corresponding to the minimum three-body coupling required for the emergence of a robust hysteresis loop.

The phase diagrams reveal the same qualitative behavior observed for SIS dynamics. In the absence of nestedness [$\alpha_{1,2}=0$, panel (a)], the synchronization threshold is nearly independent of the higher-order coupling, apart from the small finite-size fluctuations close to the transition, while a broad bistable region develops for sufficiently large values of $\sigma_2$. As nestedness increases, the bistable region progressively shrinks and the synchronization threshold becomes increasingly sensitive to the three-body coupling, indicating that nestedness amplifies the effect of higher-order interactions on the onset of synchronization. The inset in panel (e) highlights the small residual bistable region emerging immediately above $\hat{\sigma}_2$.

The same behavior is illustrated in panels (b,d,f), which report representative forward and backward continuations at fixed $\sigma_2=3$. The separation between the two branches progressively decreases with increasing nestedness, eventually leading to a continuous transition for $\alpha_{1,2}=1$. This trend is quantified in panel (h), where the normalized bistable area $A_{\rm BI}/A_{\rm tot}$ decreases monotonically with $\alpha_{1,2}$.

Finally, panel (g) reports the synchronization threshold $\sigma_1^*$ as a function of the higher-order coupling $\sigma_2$ for all values of $\alpha_{1,2}$. Although the numerically extracted curves exhibit visible finite-size fluctuations, owing to the smaller system size employed here ($N=600$) and the substantially higher computational cost of integrating higher-order Kuramoto dynamics, the overall trend is clear. Consistently with the analytical prediction obtained for SIS dynamics, the synchronization threshold is almost independent of the higher-order coupling for $\alpha_{1,2}=0$, whereas increasing nestedness progressively strengthens its dependence on $\sigma_2$. Overall, these results confirm that the same microscopic mechanism identified for SIS spreading also governs synchronization dynamics with three-body interactions.

\begin{figure}
    \centering
    \includegraphics[width=\linewidth]{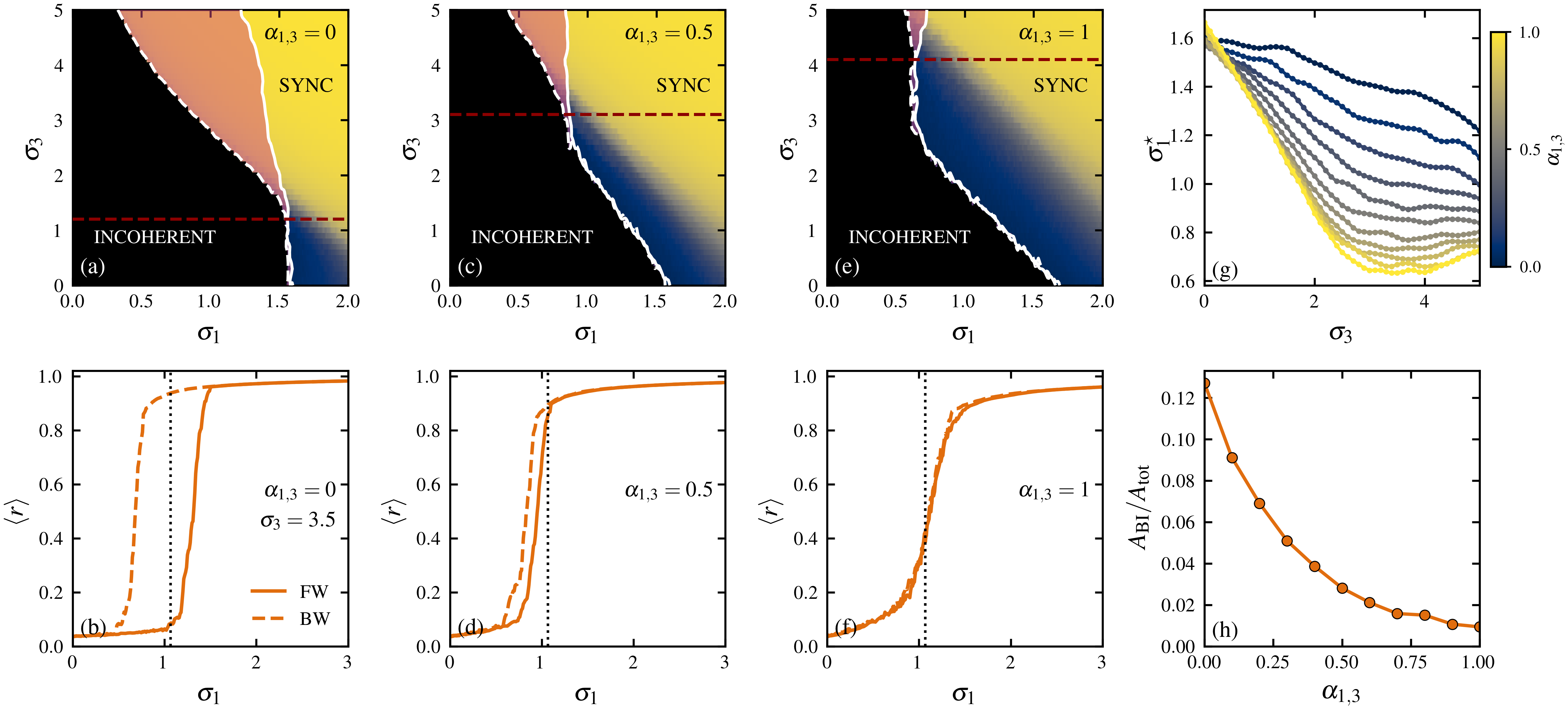}
\caption{
\textbf{Effect of nestedness on Kuramoto dynamics with pairwise and four-body interactions.}
The same analysis reported in Supplementary Figure~\ref{fig:sm_kuramoto_12} is repeated for systems with pairwise and four-body interactions only. Panels (a,c,e) show the phase diagrams in the $(\sigma_1,\sigma_3)$ plane for three representative values of the nestedness parameter $\alpha_{1,3}$. The white solid line denotes the synchronization onset obtained from the forward branch, while the white dashed line marks the backward transition delimiting the bistable region. The red dashed horizontal lines indicate the numerically estimated critical higher-order coupling $\hat{\sigma}_3$. Panels (b,d,f) report the corresponding forward (solid) and backward (dashed) stationary synchronization order parameter $r^\star$ as a function of $\sigma_1$, and the vertical dotted lines denote the numerically extracted critical value $\sigma_1^\star$ for $\alpha_{1,3}=0$. Panel (g) shows the dependence of $\sigma_1^\star$ on the four-body coupling $\sigma_3$ for different values of $\alpha_{1,3}$, while panel (h) reports the normalized bistable area $A_{\mathrm{BI}}/A_{\mathrm{tot}}$ as a function of nestedness. As in the pairwise--three-body case, increasing nestedness simultaneously anticipates the onset of synchronization and suppresses the bistable region, showing that the effect is robust to both the interaction order and the size of the higher-order groups.
}  \label{fig:sm_kuramoto_13}
\end{figure}

We next consider case (II), where pairwise interactions coexist only with four-body hyperedges. We generate random regular hypergraphs with $N=600$, $k_1=7$, and $k_3=2$, while the set of three-body interactions is empty. As before, we construct $11$ hypergraphs with tunable nestedness $\alpha_{1,3}$ and compute the corresponding synchronization diagrams in the $(\sigma_1,\sigma_3)$ plane.

The results, shown in Supplementary Figure~\ref{fig:sm_kuramoto_13}, closely mirror those obtained for pairwise and three-body interactions. The phase diagrams in panels (a,c,e) show that increasing $\alpha_{1,3}$ progressively reduces the bistable region and strengthens the dependence of the synchronization threshold on the four-body coupling. The representative continuations at fixed $\sigma_3=3.5$ in panels (b,d,f) confirm the gradual suppression of hysteresis, while panel (h) shows the corresponding decrease of $A_{\rm BI}/A_{\rm tot}$.

As in the $\alpha_{1,2}$ case, panel (g) reports $\sigma_1^*$ as a function of the higher-order coupling. Despite the finite-size effects, increasing nestedness systematically enhances the influence of four-body interactions on the synchronization onset while suppressing the bistable regime. 

To verify that the observed phenomenology is not specific to the higher-order coupling proposed in Ref.~\cite{skardal2020higher}, we finally consider an alternative model for three-body interactions. For simplicity, we restrict the analysis to systems with pairwise and three-body interactions only, and study the same family of random regular hypergraphs with tunable nestedness $\alpha_{1,2}$ considered above.

The dynamics is described by
\begin{equation}
\dot{\theta}_i
=
\omega_i
+
\frac{\sigma_1}{k_1}
\sum_{j=1}^{N}
a_{ij}^{(1)}
\sin(\theta_j-\theta_i)
+
\frac{\sigma_2}{2k_2}
\sum_{j,k=1}^{N}
a_{ijk}^{(2)}
\sin(\theta_j+\theta_k-2\theta_i),
\label{eq:kuramoto_alternative_SM}
\end{equation}
where the notation follows that introduced above. Synchronization is again quantified through the Kuramoto order parameter defined in Eq.~\eqref{eq:kuramoto_order_parameter}. All numerical simulations are performed using the same integration scheme, continuation protocol, and extraction procedure for the critical couplings described in the previous section.

Therefore, we consider the same family of regular random hypergraphs with $N=300$, $k_1=5$, and $k_2=2$, and tunable nestedness $\alpha_{1,2}$. For each parameter set $(\alpha_{1,2},\sigma_1,\sigma_2)$, we perform $200$ independent simulations. Again, half of the realizations are initialized with phases narrowly distributed around the synchronized state, while the remaining half start from phases uniformly distributed in $[0,2\pi]$, providing the forward and backward branches, respectively. For each parameter set, the stationary Kuramoto order parameter is computed for every realization and the average is used to construct the corresponding synchronization branches.

The results are reported in Supplementary Figure~\ref{fig:sm_kuramoto_alternative}. Panel (a) summarizes the numerically estimated critical pairwise coupling $\sigma_1^*$ together with the critical higher-order coupling $\hat{\sigma}_2$ as functions of the nestedness parameter $\alpha_{1,2}$. Panels (b--d) show representative forward and backward synchronization diagrams for $\sigma_2=3$ and three representative values of $\alpha_{1,2}$. As in the model presented in Eqs.\eqref{eq:kuramoto_skardal}, increasing nestedness progressively anticipates the synchronization transition while simultaneously suppressing the hysteresis loop. For fully nested structures ($\alpha_{1,2}=1$) the transition is continuous and the forward and backward branches nearly coincide, whereas decreasing nestedness enlarges the bistable region, eventually leading to pronounced explosive synchronization for $\alpha_{1,2}=0$.

\begin{figure}[t!]
    \centering
    \includegraphics[width=0.9\linewidth]{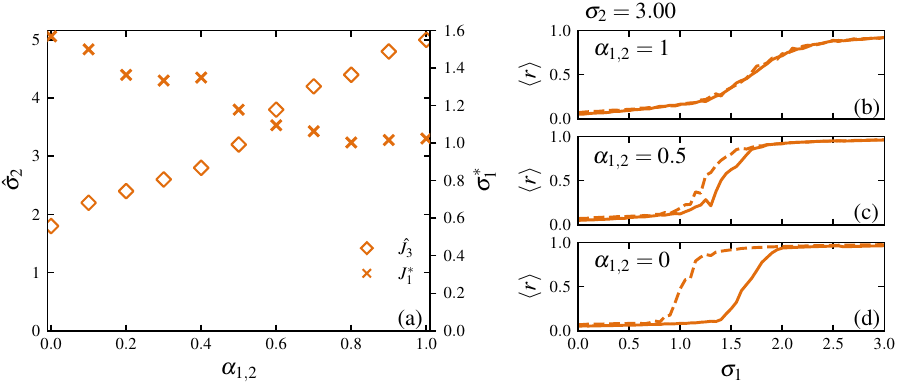}
\caption{
\textbf{Robustness of the effect of nestedness to an alternative higher-order coupling function.}
(a) Numerically estimated critical pairwise coupling $\sigma_1^*$ (crosses) and critical three-body coupling $\hat{\sigma}_2$ (diamonds) as functions of the nestedness parameter $\alpha_{1,2}$, obtained from simulations on regular hypergraphs with $N=300$, $k_1=5$, and $k_2=2$. Panels (b--d) show the time-averaged Kuramoto order parameter $\langle r\rangle$ as a function of $\sigma_1$ for fixed $\sigma_2=3$ and $\alpha_{1,2}=1$, $0.5$, and $0$, respectively. Solid and dashed curves correspond to forward and backward continuations. Increasing nestedness shifts the synchronization onset towards lower values of $\sigma_1$ while progressively reducing the hysteresis region, confirming that the dual effect of nestedness persists for a different microscopic form of the three-body coupling.
}
    \label{fig:sm_kuramoto_alternative}
\end{figure}
The analyses reported in this Supplemental Material show that the effect of nestedness extends beyond spreading dynamics and beyond systems with pairwise and three-body interactions. This indicates that the dual effect is determined by the microscopic organization of interactions across different group sizes, rather than emerging as a process-specific feature of the dynamics under consideration.

\section{Higher-order networks with tunable nestedness}
\label{sec:rewiring_procedure}
Throughout this work we consider regular higher-order networks with interactions up to order $M=3$, allowing for pairwise interactions (1-hyperedges), three-body interactions (2-hyperedges), and four-body interactions (3-hyperedges). Each node belongs to exactly $k_1$, $k_2$, and $k_3$ hyperedges of the corresponding order. Depending on the dynamical process under consideration, we study three different type of nestedness, namely between pairwise and three-body interactions ($\alpha_{1,2}$), pairwise and four-body interactions ($\alpha_{1,3}$), and three-body and four-body interactions ($\alpha_{2,3}$). We describe here the numerical procedure used to generate ensembles of regular hypergraphs with prescribed values of these nestedness parameters.

For a given pair of interaction orders $(m,n)$ with $m<n$, we first generate the set of $n$-hyperedges using configuration-model-like algorithms for regular hypergraphs, rejecting realizations containing repeated nodes within the same hyperedge or duplicated hyperedges. Once generated, the higher-order interaction set is kept fixed throughout the entire construction.

We then construct the lower-order interaction set so as to obtain the maximally nested configuration, corresponding to $\alpha_{m,n}=1$. To this end, lower-order interactions are selected exclusively among those contained within the higher-order hyperedges, while preserving the prescribed regular degree of every node. The resulting hypergraph therefore maximizes the number of lower-order interactions embedded within higher-order ones under the imposed degree constraints.

Starting from this fully nested configuration, lower values of $\alpha_{m,n}$ are obtained through a degree-preserving rewiring procedure applied only to the lower-order interaction set, while the higher-order hyperedges remain unchanged. At each rewiring step, two lower-order interactions are selected uniformly at random and one node is chosen from each of them. The selected nodes are exchanged, producing two candidate interactions. The proposed rewiring is accepted only if it preserves the regular degree of every node, does not generate self-loops or duplicated interactions, and does not increase the target nestedness. Otherwise, the original configuration is restored. By iterating this procedure, the fraction of lower-order interactions embedded within higher-order ones is progressively reduced, allowing the complete interval $\alpha_{m,n}\in[0,1]$ to be explored while preserving all degree constraints.

This construction generates ensembles of regular higher-order networks with identical interaction degrees and identical higher-order interaction sets, differing only in the amount of nestedness between the selected interaction orders. Consequently, all changes observed in the dynamical behavior throughout the manuscript can be directly attributed to nestedness, independently of degree heterogeneity or other structural properties of the underlying hypergraph.

\section{Closures for the SIS homogeneous mean-field model with nestedness and disentangling by interaction order}
\label{sec:clossures_SM}
In this section, we present the closures used to approximate the system introduced in the main text to model SIS dynamics on regular hypergraphs. As mentioned in the End Matter, we consider two interaction orders: pairwise (1-hyperedges) and three-body (2-hyperedges). Susceptible nodes become infected either through a link at rate $\beta_1$ or through a 2-hyperedge containing two infected neighbors at rate $\beta_2$, while infected nodes recover at rate $\mu$. The microscopic nestedness between the two interaction orders is controlled by the inter-order overlap $\alpha_{1,2}$. Throughout the remainder of this Supplemental Material, we use the shorthand $\alpha\equiv\alpha_{1,2}$.

We track the density variables $\rho^{\rm I},\rho^{\rm SI},\rho^{\rm SSS_\Delta},\rho^{\rm SSI_\Delta}$ and $\rho^{\rm ISI_\Delta}$, where $\rho^{\rm A}=[A]/N$, $\rho^{\rm AB}=[AB]/(Nk_1)$, and $\rho^{\rm ABC_\Delta}=[ABC_\Delta]/(2Nk_2)$ denote normalized motif densities (node, link, and 2-hyperedge states, respectively). By homogeneity and permutation symmetry, $\rho^{\rm SI}=\rho^{\rm IS}$ and, e.g., $\rho^{\rm ISI_\Delta}=\rho^{\rm IIS_\Delta}=\rho^{\rm SII_\Delta}$. To close the resulting hierarchy we additionally require densities for higher-order composite motifs, which we define as follows:
\begin{itemize}
   \item \textbf{Composite motifs of two pairwise interactions}: $\rho^{\rm ABC}$ denotes the density of pairs of links sharing a common node, where the shared node is in state $\rm B$ and the two other nodes are in states $\rm A$ and $\rm C$.
    \item \textbf{Composite motifs of a pairwise and a three-body interaction}: $\rho^{\rm ABC_{\Delta}D}$ denotes the density of 2-hyperedges whose nodes are in states $\rm A$, $\rm B$, and $\rm C$, with the additional constraint that the node in state $\rm C$ also participates in a link with a node in state $\rm D$.
    \item \textbf{Composite motifs of two three-body interactions}: $\rho^{\rm ABCDE_{\bowtie}}$ denotes the density of 2-hyperedges whose nodes are in states $\rm A$, $\rm B$, and $\rm C$, where the node in state $\rm C$ also belongs to a second 2-hyperedge with nodes in states $\rm D$ and $\rm E$.
\end{itemize}

The dynamics of the state-vector variables is then governed by the following system of differential equations:
\begin{equation}
\label{eq:model_system_eqs}
    \begin{array}{ll}
         \displaystyle \dot{\rho}^{\rm I} &= -\mu \rho^{\rm I} + \beta_1 k_1 \rho^{\rm SI} + \beta_2 k_2 \rho^{\rm ISI_\Delta}; \\[10pt]
\dot{\rho}^{\rm SI} &= \beta_1 \left[ \left( k_1 - 1\right) \left(\rho^{\rm SSI} - \rho^{\rm ISI}\right) - \rho^{\rm SI}  \right]    + \beta_2\frac{k_2}{k_1} \left[\left(k_1 - 2\alpha\right) \left(\rho^{\rm IIS_\Delta S} -\rho^{\rm IIS_\Delta I} \right) -2 \alpha \rho^{\rm ISI_\Delta} \right]+ \mu \left(\rho^{\rm II} - \rho^{\rm SI}\right); \\[10pt]
\dot{\rho}^{\rm SSS_\Delta} &= -3\beta_1\left(k_1-2\alpha \right) \rho^{\rm SSS_\Delta I} - 3\beta_2\left(k_2-1 \right)\rho^{\rm SSSII_{\bowtie}}  + 3\mu\rho^{\rm SSI_\Delta}; \\[10pt]
    \dot{\rho}^{\rm SSI_\Delta} &= \beta_1 \left[\left(k_1 -2\alpha\right) \left( \rho^{\rm SSS_\Delta I} -2 \rho^{\rm ISS_\Delta I}\right) - 2\alpha \rho^{\rm SSI_\Delta}\right]  + \beta_2 \left(k_2 -1\right) \left( \rho^{\rm SSSII_{_{\bowtie}}} - 2\rho^{\rm ISSII_{\bowtie}} \right)+ \\ [5pt] &\mu \left( 2\rho^{\rm ISI_\Delta} - \rho^{\rm SSI_\Delta}\right); 
    \\[10pt]
         \dot{\rho}^{\rm ISI_\Delta} &= \beta_1 \left[\left(k_1-2\alpha\right) \left(2 \rho^{\rm ISS_\Delta I } - \rho^{\rm IIS_\Delta I} \right)\right] + 2\alpha \beta_1 \left( \rho^{\rm SSI_\Delta} - \rho^{\rm ISI_\Delta}\right) +\beta_2 \left[\left( k_2 - 1\right) \left( 2 \rho^{\rm ISSII_{\bowtie}} - \rho^{\rm IISII_{\bowtie }}\right) - \rho^{\rm ISI_\Delta} \right] \\ [5pt] &
          + \mu \left(\rho^{\rm III_\Delta} - 2 \rho^{\rm ISI_\Delta}\right). 
    \end{array}
\end{equation}

This system is exact, but it is not closed: it depends on the evolution of composite motifs that are not themselves included among the state variables. To obtain a solvable system, we close the hierarchy at the level of these composite motifs. Specifically, we neglect (i) closed triples of links, i.e., triangular clustering within the pairwise layer, and (ii) intra-order overlap among distinct 2-hyperedges (see also Ref.~\cite{burgio2024triadic}). Under these two assumptions, each higher-order composite motif factorizes into a product of lower-order densities, giving
\begin{equation}
\label{eq:system_closures}
\begin{aligned}
\rho^{\rm ISI} &= \frac{(\rho^{\rm SI})^2}{\rho^{\rm S}}, 
&\qquad
\rho^{\rm SSI} &= \frac{\rho^{\rm SS}\rho^{\rm SI}}{\rho^{\rm S}}, &\qquad
\rho^{\rm SSS_\Delta I} &= \frac{\rho^{\rm SSS_\Delta}\rho^{\rm SI}}{\rho^{\rm S}}, 
\\[3pt]
\rho^{\rm ISS_\Delta I} &= \frac{\rho^{\rm SSI_\Delta}\rho^{\rm SI}}{\rho^{\rm S}}, &\qquad
\rho^{\rm IIS_\Delta S} &= \frac{\rho^{\rm ISI_\Delta}\rho^{\rm SS}}{\rho^{\rm S}}, 
&\qquad
\rho^{\rm IIS_\Delta I} &= \frac{\rho^{\rm ISI_\Delta}\rho^{\rm SI}}{\rho^{\rm S}}, \\[3pt]
\rho^{\rm ISSII_{\bowtie}} &= \frac{\rho^{\rm SSI_\Delta}\rho^{\rm ISI_\Delta}}{\rho^{\rm S}}, 
&\qquad
\rho^{\rm SSSII_{\bowtie}} &= \frac{\rho^{\rm SSS_\Delta}\rho^{\rm ISI_\Delta}}{\rho^{\rm S}}, &\qquad
\rho^{\rm IISII_{\bowtie}} &= \frac{(\rho^{\rm ISI_\Delta})^2}{\rho^{\rm S}}.
\end{aligned}
\end{equation}

Substituting these closures into Eqs.~\eqref{eq:model_system_eqs} yields the final closed system of equations used throughout the main text:
\begin{equation}
\label{eq:closed_system}
    \begin{array}{ll}
         \displaystyle \dot{\rho}^{\rm I} &= -\mu \rho^{\rm I} + \beta_1 k_1 \rho^{\rm SI} + \beta_2 k_2 \rho^{\rm ISI_\Delta}; \\[10pt]
\dot{\rho}^{\rm SI} &= \beta_1 \left[ \left( k_1 - 1\right) \frac{\rho^{\rm SI}}{\rho^{\rm S}}\left(\rho^{\rm SS} - \rho^{\rm SI}\right) - \rho^{\rm SI}  \right]   + \beta_2\frac{k_2}{k_1} \left[\left(k_1 - 2\alpha\right) \frac{\rho^{\rm ISI_\Delta}}{\rho^{\rm S}}\left(\rho^{\rm SS} - \rho^{\rm SI}\right) -2 \alpha \rho^{\rm ISI_\Delta} \right]  + \mu \left(\rho^{\rm II} - \rho^{\rm SI}\right); \\[10pt]
\dot{\rho}^{\rm SSS_\Delta} &= -3\frac{\rho^{\rm SSS_\Delta}}{\rho^{\rm S}}\left[\beta_1\left(k_1-2\alpha \right) \rho^{\rm SI} + \beta_2\left(k_2-1 \right)\rho^{\rm ISI_\Delta}\right]  + 3\mu\rho^{\rm SSI_\Delta}; \\[10pt]
    \dot{\rho}^{\rm SSI_\Delta} &= \beta_1 \left[\left(k_1 -2\alpha\right) \frac{\rho^{\rm SI}}{\rho^{\rm S}}\left( \rho^{\rm SSS_\Delta} -2 \rho^{\rm SSI_\Delta}\right) - 2\alpha \rho^{\rm SSI_\Delta}\right]  + \beta_2 \left(k_2 -1\right) \frac{\rho^{\rm ISI_\Delta}}{\rho^{\rm S}}\left( \rho^{\rm SSS_\Delta} - 2\rho^{\rm SSI_\Delta} \right) \\[5pt] &+ \mu \left( 2\rho^{\rm ISI_\Delta} - \rho^{\rm SSI_\Delta}\right); 
    \\[10pt]
         \dot{\rho}^{\rm ISI_\Delta} &= \beta_1 \left(k_1-2\alpha\right) \frac{\rho^{\rm SI}}{\rho^{\rm S}}\left(2 \rho^{\rm SSI_\Delta} - \rho^{\rm ISI_\Delta} \right)  + 2\alpha \beta_1 \left( \rho^{\rm SSI_\Delta} - \rho^{\rm ISI_\Delta}\right) \\[5pt] & +\beta_2 \left[\left( k_2 - 1\right) \frac{\rho^{\rm ISI_\Delta}}{\rho^{\rm S}}\left( 2 \rho^{\rm SSI_\Delta} - \rho^{\rm ISI_\Delta}\right) - \rho^{\rm ISI_\Delta} \right] 
          + \mu \left(\rho^{\rm III_\Delta} - 2 \rho^{\rm ISI_\Delta}\right)
         . \\[10pt]      
    \end{array}
\end{equation}
 
The remaining densities can be obtained from the conservation identities:
\begin{equation}
\begin{aligned}
        \rho^{\rm S} &= 1-\rho^{\rm I},\\
    \rho^{\rm II} &= \rho^{\rm I}-\rho^{\rm SI}, \\
    \rho^{\rm SS} &= \rho^{\rm S}-\rho^{\rm SI}, \\
    \rho^{\rm III_\Delta} &= 1-\rho^{\rm SSS_\Delta}-3\rho^{\rm SSI_\Delta}-3\rho^{\rm ISI_\Delta}.
\end{aligned}
\label{eq:conservation_identities}
\end{equation}

\section{Disentangling the SIS homogeneous mean-field model by interaction order and internal/external channels}

Equations~\eqref{eq:model_system_eqs}  with the closures in \eqref{eq:system_closures} already provide a closed description of the system. However, they aggregate contributions from microscopically distinct transmission processes into a single rate equation for each motif density, obscuring the individual role played by three-body infection versus pairwise infection, and, within the latter, by pairwise infection occurring inside versus outside a group. To isolate these mechanisms, we disentangle Eqs.~\eqref{eq:model_system_eqs} by decomposing each motif density into the sum of the contributions that generate it: a genuine three-body infection term (subscript $2$) and a link-driven infection term, which for group-state densities is further split into transmission occurring on links internal to the group and on links external to it (subscripts $1,\mathrm{int}$ and $1,\mathrm{ext}$, respectively). This decomposition is exact, since it simply partitions the right-hand side of each equation in Eqs.~\eqref{eq:model_system_eqs} according to the microscopic event responsible for each term, without introducing any further approximation.

Specifically, for the case of $\rho^{SI}$ we can decompose its differential equation into a pairwise contribution $\rho^{SI}_1$ and a three-body contribution $\rho^{SI}_2$ as:
\begin{equation}
\begin{aligned}{}
     \dot{\rho}^{\rm SI}_1 &= \beta_1 \left[ \left( k_1 - 1\right) \left(\rho^{\rm SSI} - \rho^{\rm ISI}\right) - \rho^{\rm SI}  \right]  + \mu \left(\rho^{\rm II}_1 - \rho^{\rm SI}_1\right); \\[5pt] 
     \dot{\rho}^{\rm SI}_2 &= \beta_2\frac{k_2}{k_1} \left[\left(k_1 - 2\alpha\right) \left(\rho^{\rm IIS_\Delta S} -\rho^{\rm IIS_\Delta I} \right) -2 \alpha \rho^{\rm ISI_\Delta} \right] + \mu \left(\rho^{\rm II}_2 - \rho^{\rm SI}_2\right).
\end{aligned}
\end{equation}

A similar decomposition can be applied to $\rho^{II}$ such that $\rho^{II} = \rho_1^{II}+\rho_2^{II}$, which, based on the conservation identities \eqref{eq:conservation_identities}, can be expressed in terms of the decomposition of $\rho^{SI}$ and $\rho^{I}$. This gives:
\begin{align}
    \rho_1^{II} &= \rho_1^{I} -\rho_1^{SI}; \\
    \rho_2^{II} &= \rho_2^{I} -\rho_2^{SI}.
\end{align}

For the group-state densities, we can further separate the pairwise contribution into external and internal channels, considering those terms proportional to $\alpha$ as arising from contributions that include possible overlapping (nested) interactions. In this way, the group-state variables can be disentangled as:
\begin{equation}
\label{eq:groups_disentangled_int_ext}
    \begin{array}{ll}
\dot{\rho}^{\rm SSS_\Delta}_{1,\rm ext} =& -3\beta_1\left(k_1-2\alpha \right) \rho^{\rm SSS_\Delta I} + 3\mu\rho^{\rm SSI_\Delta}_{1,\rm ext}; \\[5pt]
\dot{\rho}^{\rm SSS_\Delta}_{1,\rm int} =&  + 3\mu\rho^{\rm SSI_\Delta}_{1,\rm int};  \\[5pt]
\dot{\rho}^{\rm SSS_\Delta}_{2} =& - 3\beta_2\left(k_2-1 \right)\rho^{\rm SSSII_{\bowtie}} + 3\mu\rho^{\rm SSI_\Delta}_{2}; \\[5pt]
    \dot{\rho}^{\rm SSI_\Delta}_{1,\rm ext} =& \beta_1 \left[\left(k_1 -2\alpha\right) \left( \rho^{\rm SSS_\Delta I} -2 \rho^{\rm ISS_\Delta I}\right) \right] + \mu \left( 2\rho^{\rm ISI_\Delta}_{1,\rm ext} - \rho^{\rm SSI_\Delta}_{1,\rm ext}\right); 
    \\ [5pt]
        \dot{\rho}^{\rm SSI_\Delta}_{1,\rm int} =& -2 \beta_1\alpha \rho^{\rm SSI_\Delta} + \mu \left( 2\rho^{\rm ISI_\Delta}_{1,\rm int} - \rho^{\rm SSI_\Delta}_{1,\rm int}\right); 
    \\ [5pt]
        \dot{\rho}^{\rm SSI_\Delta}_2 =& \beta_2 \left(k_2 -1\right) \left( \rho^{\rm SSSII_{_{\bowtie}}} - 2\rho^{\rm ISSII_{\bowtie}} \right) + \mu \left( 2\rho^{\rm ISI_\Delta}_2 - \rho^{\rm SSI_\Delta}_2\right); 
    \\ [5pt]
         \dot{\rho}^{\rm ISI_\Delta}_{1,\rm ext} =& + \mu \left( \rho^{\rm III_\Delta}_{1,\rm ext} - 2 \rho^{\rm ISI_\Delta}_{1,\rm ext}\right) + \beta_1 \left( k_1 - 2\alpha \right) \left ( 2\rho^{\rm ISS_\Delta I }- \rho^{\rm IIS_\Delta I }\right);  \\ [5pt]
             \dot{\rho}^{\rm ISI_\Delta}_{1,\rm int} =& + \mu \left( \rho^{\rm III_\Delta}_{1,\rm int} - 2 \rho^{\rm ISI_\Delta}_{1,\rm int}\right) + 2 \alpha \beta_1 \left( \rho^{\rm SSI_\Delta} - 2 \rho^{\rm ISI_\Delta}\right); \\[5pt]
             \dot{\rho}^{\rm ISI_\Delta}_{2} =&  + \mu \left( \rho^{\rm III_\Delta}_{2} - 2 \rho^{\rm ISI_\Delta}_{2}\right)  +\beta_2 \left[\left( k_2 - 1\right) \left( 2 \rho^{\rm ISSII_{\bowtie}} - \rho^{\rm IISII_{\bowtie }}\right) - \rho^{\rm ISI_\Delta} \right].
    \end{array}
\end{equation}

\section{Derivation of the fast variables and disentangling by interaction order}

In the early stage ($\rho^{\rm I}\to 0$), ratios of motif densities, such as $\rho^{\rm SI}/\rho^{\rm I}$, relax on a fast time scale. We define these ratios as fast variables, using the notation $\Pi\equiv\rho^{\rm SI}/\rho^{\rm I}$ and $\Psi\equiv\rho^{\rm ISI_\Delta}/\rho^{\rm I}$, so that $\delta\equiv\rho^{\rm ISI_\Delta}/\rho^{\rm SI}=\Psi/\Pi$ (main text).
To close the fast-variable dynamics we introduce two additional ratios, $\Omega\equiv\rho^{\rm SSS_\Delta}/\rho^{\rm I}$ and
$\Upsilon\equiv\rho^{\rm SSI_\Delta}/\rho^{\rm I}$.
The remaining pair fast variable is eliminated using the identity
$\rho^{\rm SI}+\rho^{\rm II}=\rho^{\rm I}$, i.e., $\rho^{\rm II}/\rho^{\rm I}=1-\Pi$.
Fast-variable equations follow by differentiating these ratios via the chain rule. We illustrate the derivation explicitly for $\Pi$; the remaining fast variables $\Upsilon,\Psi,\Omega$ follow analogously. By the chain rule,
\begin{equation}
    \dot{\Pi}=\frac{\dot{\rho}^{\rm SI}}{\rho^{\rm I}}-\Pi\frac{\dot{\rho}^{\rm I}}{\rho^{\rm I}}.
\end{equation}

Using the equations for $\dot{\rho}^{\rm SI}$ and $\dot{\rho}^{\rm I}$ from Eqs.~\eqref{eq:closed_system}, we substitute and obtain:

\begin{equation}
\begin{aligned}
    \dot{\Pi}=&\frac{1}{\rho^{\rm I}}\bigg( \beta_1 \left[ \left( k_1 - 1\right) \frac{\rho^{\rm SI}}{\rho^{\rm S}}\left(\rho^{\rm SS} - \rho^{\rm SI}\right) - \rho^{\rm SI}  \right]   + \beta_2\frac{k_2}{k_1} \left[\left(k_1 - 2\alpha\right) \frac{\rho^{\rm ISI_\Delta}}{\rho^{\rm S}}\left(\rho^{\rm SS} - \rho^{\rm SI}\right) -2 \alpha \rho^{\rm ISI_\Delta} \right]  + \mu \left(\rho^{\rm II} - \rho^{\rm SI}\right) \bigg)\\&-\frac{\Pi}{\rho^{\rm I}}\bigg( -\mu\rho^{\rm I}+\beta_1 k_1\rho^{\rm SI}+\beta_2 k_2\rho^{\rm ISI_\Delta}\bigg).
    \end{aligned}
\end{equation}
Simplifying, we can write
\begin{equation}
\label{eq:fast_pi_ratios}
\begin{aligned}
    \dot{\Pi}=&\bigg( \beta_1 \left[ \left( k_1 - 1\right) \Big(\frac{\rho^{\rm SI}\rho^{\rm SS}}{\rho^{\rm S}\rho^{\rm I}} - \frac{(\rho^{\rm SI})^2}{\rho^{\rm S}\rho^{\rm I}}\Big) -\frac{\rho^{\rm SI}}{\rho^{\rm I}}  \right] 
    + \beta_2\frac{k_2}{k_1} \left[\left(k_1 - 2\alpha\right) \Big(\frac{\rho^{\rm ISI_\Delta}\rho^{\rm SS}}{\rho^{\rm S} \rho^{\rm I}} - \frac{\rho^{\rm ISI_\Delta}\rho^{\rm SI}}{\rho^{\rm S} \rho^{\rm I}}\Big) -2 \alpha \frac{\rho^{\rm ISI_\Delta}}{\rho^{\rm I}} \right] 
    + \mu \left(\frac{\rho^{\rm II}}{\rho^{\rm I}} - \frac{\rho^{\rm SI}}{\rho^{\rm I}}\right) \bigg)
    \\&-\Pi\bigg( -\mu+\beta_1 k_1\frac{\rho^{\rm SI}}{\rho^{\rm I}}+\beta_2 k_2\frac{\rho^{\rm ISI_{\Delta}}}{\rho^{\rm I}}\bigg).
    \end{aligned}
\end{equation}

As we are considering the early stage of the contagion process, $(\rho^{\rm I},\rho^{\rm S},\rho^{\rm SS},\rho^{\rm SI}, \rho^{\rm II}, \rho^{\rm SSS_{\Delta}},\rho^{\rm SSI_{\Delta}},\rho^{\rm ISI_{\Delta}}) \to (0,1,1,0,0,1,0,0)$. This implies that some of the ratios in Eq.~\eqref{eq:fast_pi_ratios} vanish, with the remaining terms given by:
\begin{equation}
\label{eq:fast_pi_leading}
\begin{aligned}
    \dot{\Pi}=&\bigg( \beta_1 \left[ \left( k_1 - 1\right) \frac{\rho^{\rm SI}}{\rho^{\rm I}}  -\frac{\rho^{\rm SI}}{\rho^{\rm I}}  \right] 
    + \beta_2\frac{k_2}{k_1} \left[\left(k_1 - 2\alpha\right) \frac{\rho^{\rm ISI_\Delta}}{\rho^{\rm I}} -2 \alpha \frac{\rho^{\rm ISI_\Delta}}{\rho^{\rm I}} \right] 
    + \mu \left(\frac{\rho^{\rm II}}{\rho^{\rm I}} - \frac{\rho^{\rm SI}}{\rho^{\rm I}}\right) \bigg)
    \\&-\Pi\bigg( -\mu+\beta_1 k_1\frac{\rho^{\rm SI}}{\rho^{\rm I}}+\beta_2 k_2\frac{\rho^{\rm ISI_{\Delta}}}{\rho^{\rm I}}\bigg).
    \end{aligned}
\end{equation}

Substituting the remaining ratios of densities with the fast-variable notation, we can write the equation as:
\begin{equation}
\label{eq:fast_pi_fastvars}
\begin{aligned}
    \dot{\Pi}=&\bigg( \beta_1 \left( k_1 - 2\right) \Pi 
    + \beta_2\frac{k_2}{k_1}\left(k_1 - 4\alpha\right) \Psi
    + \mu \left(1-2\Pi\right) \bigg)
    -\Pi\bigg( -\mu+\beta_1 k_1\Pi+\beta_2 k_2\Psi\bigg),
    \end{aligned}
\end{equation}
which can be rearranged as:
\begin{equation}
\label{eq:fast_pi_final}
\begin{aligned}
    \dot{\Pi}=  \mu (1-\Pi) + \beta_1  \left( k_1 - 2\right) \Pi 
    + \beta_2\frac{k_2}{k_1} \left(k_1 - 4\alpha\right) \Psi - \beta_1 k_1\Pi^2 - \beta_2 k_2\Pi\Psi.
    \end{aligned}
\end{equation}

Proceeding similarly for $\Upsilon,\Psi,\Omega$, we obtain the full closed system of fast-variable equations:
\begin{equation}
\label{eq:fastvars_system}
\begin{aligned}
\dot{\Pi} =& \mu(1-\Pi) + \beta_1\left(k_1-2\right) \Pi
          + \beta_2\frac{k_2}{k_1}(k_1-4\alpha)\Psi - \beta_1 k_1 \Pi^2 - \beta_2 k_2 \Pi\Psi,\\
\dot{\Upsilon} =& \beta_1(k_1-2\alpha)\Pi - 2\beta_1\alpha\,\Upsilon
             + \big(2\mu+\beta_2(k_2-1)\big)\Psi 
             - \beta_1 k_1 \Pi\Upsilon - \beta_2 k_2 \Upsilon\Psi,\\
\dot{\Omega} =& \mu (2\Upsilon+\Omega)-3\beta_1(k_1 - 2\alpha)\Pi - 3\beta_2(k_2-1)\Psi 
            - \beta_1  k_1 \Pi \Omega -\beta_2 k_2 \Omega \Psi,\\
\dot{\Psi} =& 2\beta_1\alpha\,\Upsilon +\big(-2\beta_1\alpha-\mu-\beta_2\big)\Psi +\mu\,\big(1-\Omega
            -3\Upsilon-3\Psi\big)- \beta_1 k_1 \Pi\Psi - \beta_2 k_2  \Psi^2.
\end{aligned}
\end{equation}
This system directly provides the early-time evolution of $\Psi=\rho^{\rm ISI_\Delta}/\rho^{\rm I}$ and hence of
$\delta=\Psi/\Pi=\rho^{\rm ISI_\Delta}/\rho^{\rm SI}$.

Furthermore, the fast variables can be decomposed in a similar way as we did for the original system of equation in the previous section. The variable $\Pi$, which depends on $\rho^{SI}$ and $\rho^{I}$, can be disentangled into pairwise and three-body contributions, whereas $\Upsilon$, $\Psi$, and $\Omega$ have their pairwise contribution further decomposed into internal and external components:
\begin{equation}
\begin{array}{l}
\Pi=\Pi_1 +\Pi_2;\\
\Upsilon=\Upsilon_2+\Upsilon_{1,{\rm int}}+\Upsilon_{1,{\rm ext}};\\[5pt]
\Psi=\Psi_2+\Psi_{1,{\rm int}}+\Psi_{1,{\rm ext}}; \\[5pt]
\Omega=\Omega_2+\Omega_{1,{\rm int}}+\Omega_{1,{\rm ext}}.
\end{array}
\end{equation}
For the disentangled fast variables, we cannot use the conservation identities directly, since the recovery terms depend on complementary densities that cannot themselves be separated into contributions by interaction order. For this reason, we introduce two additional densities, $\rho^{\rm II}$ and $\rho^{\rm III_\Delta}$, whose evolution equations read:

\begin{equation}
    \begin{aligned}
      \dot{\rho}^{\rm II}& = -2\mu\rho^{\rm II} + 2\beta_1 \rho^{\rm SI}  + 2\beta_1 (k_1 -1)\rho^{\rm ISI}  + 2\beta_2 \frac{k_2}{k_1} (k_1 - 2\alpha) \rho^{\rm IIS_\Delta I} + 4 \beta_2 \frac{k_2}{k_1} \alpha \rho^{\rm ISI_\Delta}, \\
         \dot{\rho}^{\rm III_{\Delta}}&= 3\beta_1\Big[\left(k_1-2\alpha\right)\rho^{\rm IIS_\Delta I} + 2\alpha \rho^{\rm ISI_\Delta} \Big] - 3\mu\rho^{\rm III_\Delta}  + 3\beta_2\Big[ \rho^{\rm ISI_\Delta} + \left(k_2-1\right) \rho^{\rm IISII_{\bowtie}}\Big].
    \end{aligned}
\end{equation}
For these two densities, we define the fast variables $\Phi \equiv \rho^{\rm II}/\rho^{\rm I}=1-\Pi$ and $\Xi \equiv \rho^{\rm III_{\Delta}}/\rho^{\rm I} = 1-\Omega-3\Upsilon-3\Psi$. Applying the chain rule as before, their evolution equations read:
\begin{equation}
    \begin{aligned}
        \dot{\Phi} &= -\mu\Phi +2\beta_1\Pi + 4\beta_2\frac{k_2}{k_1}\alpha\Psi - \beta_1 k_1 \Phi \Pi - \beta_2 k_2 \Phi \Psi,\\
        \dot{\Xi} & = -2\mu\Xi +3\beta_2\Psi +6\alpha\beta_1\Psi - \beta_1 k_1 \Xi \Pi - \beta_2 k_2 \Xi \Psi.
    \end{aligned}
\end{equation}

Denoting $\mathcal{L}(X)\equiv-\beta_1k_1\Pi\,X-\beta_2k_2\Psi\,X$, the disentangled fast-variable dynamics reads
\begin{equation}
\label{eq:fastvars_disentangled}
\begin{aligned}
\dot{\Pi}_1&= \mu\Phi_1 + \beta_1(k_1-2)\Pi  + \mathcal{L}(\Pi_1),\\
\dot{\Pi}_2&=\mu\Phi_2 + \beta_2 \frac{k_2}{k_1}\left(k_1 - 4\alpha \right)\Psi + \mathcal{L}(\Pi_2),\\
\dot{\Phi}_1 &= -\mu\Phi_1 + 2 \beta_1\Pi +\mathcal{L}(\Phi_1)\\
\dot{\Phi}_2 &= -\mu\Phi_2 + 4\beta_2\frac{k_2}{k_1}\alpha\Psi+\mathcal{L}(\Phi_2)\\\
\dot{\Upsilon}_2 &= 2\mu\,\Psi_2 + \beta_2(k_2-1)\Psi + \mathcal{L}(\Upsilon_2),\\
\dot{\Upsilon}_{1,{\rm int}} &= 2\mu\,\Psi_{1,{\rm int}} - 2\beta_1\alpha\,\Upsilon + \mathcal{L}(\Upsilon_{1,{\rm int}}),\\
\dot{\Upsilon}_{1,{\rm ext}} &= 2\mu\,\Psi_{1,{\rm ext}} + \beta_1(k_1-2\alpha)\Pi + \mathcal{L}(\Upsilon_{1,{\rm ext}}),\\[3pt]
\dot{\Psi}_2 &= \mu\,\big(\Xi_{2} - \Psi_{2}\big)   - \beta_2\Psi + \mathcal{L}(\Psi_2),\\
\dot{\Psi}_{1,{\rm int}} &= \mu\,\big(\Xi_{1,{\rm int}} - \Psi_{1,{\rm int}}\big)  + 2\alpha\beta_1(\Upsilon-\Psi) + \mathcal{L}(\Psi_{1,{\rm int}}),\\
\dot{\Psi}_{1,{\rm ext}} &= \mu\,\big(\Xi_{1,{\rm ext}} - \Psi_{1,{\rm ext}}\big) + \mathcal{L}(\Psi_{1,{\rm ext}}),\\[3pt]
\dot{\Omega}_2 &= \mu \big(2\Upsilon_2 +\Omega_2\big) - 3\beta_2(k_2-1)\Psi + \mathcal{L}(\Omega_2),\\
\dot{\Omega}_{1,{\rm int}} &= \mu \big(2\Upsilon_{1,{\rm int}} +\Omega_{1,{\rm int}} \big) + \mathcal{L}(\Omega_{1,{\rm int}}),\\
\dot{\Omega}_{1,{\rm ext}} &= \mu \big(2\Upsilon_{1,{\rm ext}} +\Omega_{1,{\rm ext}}\big) -3\beta_1(k_1-2\alpha)\Pi  + \mathcal{L}(\Omega_{1,{\rm ext}})\\
\dot{\Xi}_2 &= -2\mu\Xi_2 +3\beta_2\Psi +\mathcal{L}(\Xi_2) \\
\dot{\Xi}_{1,{\rm int}} &= -2\mu\Xi_{1,{\rm int}} +6\alpha\beta_1\Psi +\mathcal{L}(\Xi_{1,{\rm int}}) \\
\dot{\Xi}_{1,{\rm ext}} &= -2\mu\Xi_{1,{\rm ext}}  +\mathcal{L}(\Xi_{1,{\rm ext}})
\end{aligned}
\end{equation}

Lastly, since the variable $\delta=\rho^{ISI_\Delta}/\rho^{SI} = \Psi/\Pi$ is obtained from the solutions of $\rho^{ISI_\Delta}$ and $\rho^{SI}$, its disentangled decomposition is not computed via the chain rule but is instead obtained directly from the disentangled variables of $\rho^{ISI_\Delta}$ and $\rho^{SI}$:
\begin{equation}
\label{eq:fastvars_disentangled_2}
    \begin{aligned}
        \delta_2 &= \frac{\Psi_2}{\Pi},\\
        \delta_{1, \rm int} &=  \frac{\Psi_{1, \rm int}}{\Pi}, \\
        \delta_{1, \rm ext} &=  \frac{\Psi_{1, \rm ext}}{\Pi}.
    \end{aligned}
\end{equation}
Equations~\eqref{eq:fastvars_disentangled}--\eqref{eq:fastvars_disentangled_2} provide a closed early-time description of the fast variables, including the disentangled contributions by interaction order and by internal/external link channels within groups.

\section{ DERIVATION OF THE ANALYTICAL FAST VARIABLES}

We derive the quasi-stationary (early-time) values of the fast variables
$\Pi\equiv\rho^{\rm SI}/\rho^{\rm I}$ and $\delta\equiv\rho^{\rm ISI_\Delta}/\rho^{\rm SI}$
by exploiting time-scale separation close to the disease-free equilibrium.
In the regime $\rho^{\rm I}\to0$, node density $\rho^{\rm I}$ evolves slowly, while pair and group motif densities rapidly relax to values that are slaved to $\rho^{\rm I}$.
Accordingly, we treat $\rho^{\rm I}$ as quasi-constant and impose $\dot{\rho}^{\rm SI}\simeq0$ and $\dot{\rho}^{\rm ISI_\Delta}\simeq0$ (and, when needed, $\dot{\rho}^{\rm SSI_\Delta}\simeq0$), retaining only the leading-order contributions in $\rho^{\rm I}$.We start from the closed mean-field system reported in \eqref{eq:closed_system}.
Close to the disease-free state, the relevant variables scale as
$\rho^{\rm SI},\rho^{\rm SSI_\Delta},\rho^{\rm ISI_\Delta}=\mathcal{O}(\rho^{\rm I})$,
while $\rho^{\rm S}=1-\rho^{\rm I}=1+\mathcal{O}(\rho^{\rm I})$ and
$\rho^{\rm SSS_\Delta}=1+\mathcal{O}(\rho^{\rm I})$.
All composite terms are closed using the same factorized approximations of the main text, e.g.,
$\rho^{\rm ISI}\approx(\rho^{\rm SI})^2/\rho^{\rm S}$,
$\rho^{\rm SSI}\approx\rho^{\rm SS}\rho^{\rm SI}/\rho^{\rm S}$,
$\rho^{\rm IIS_\Delta I}\approx\rho^{\rm ISI_\Delta}\rho^{\rm SI}/\rho^{\rm S}$,
$\rho^{\rm IIS_\Delta S}\approx\rho^{\rm ISI_\Delta}\rho^{\rm SS}/\rho^{\rm S}$,
and similarly for the remaining composite motifs.

\emph{Step 1: quasi-stationary ratio $\bar{\delta}=\rho^{\rm ISI_\Delta}/\rho^{\rm SI}$.}
At leading order, $\dot{\rho}^{\rm ISI_\Delta}=0$ yields a linear relation between $\rho^{\rm ISI_\Delta}$ and $\rho^{\rm SI}$, because all quadratic terms (e.g., involving products of two $\mathcal{O}(\rho^{\rm I})$ quantities) can be neglected.
Using $\rho^{\rm SSI_\Delta}= \mathcal{O}(\rho^{\rm I})$ and the closure
$\rho^{\rm IIS_\Delta I}\approx\rho^{\rm ISI_\Delta}\rho^{\rm SI}$ and
$\rho^{\rm IIS_\Delta S}\approx\rho^{\rm ISI_\Delta}$ (since $\rho^{\rm SS}\simeq1$),
the dominant balance in $\dot{\rho}^{\rm ISI_\Delta}$ can be written as
\[
0 \simeq \beta_1\,\alpha k_2\lambda_1^{*2}\frac{k_1-2\alpha}{k_1}\,\rho^{\rm SI}
\;-\;\beta_2\,\alpha\lambda_1^*\lambda_2\,(k_2-1)\,\rho^{\rm ISI_\Delta}
\;-\;k_1k_2\,\rho^{\rm ISI_\Delta},
\]
where we have used the rescaled infectivities $\lambda_1=k_1\beta_1/\mu$ and $\lambda_2=k_2\beta_2/\mu$ and evaluated the expression at the epidemic threshold $\lambda_1=\lambda_1^*$.
Solving for $\rho^{\rm ISI_\Delta}/\rho^{\rm SI}$ gives the quasi-stationary value
\begin{equation}
\label{eq:delta_mf}
\bar{\delta}\equiv\frac{\rho^{\rm ISI_\Delta}}{\rho^{\rm SI}}
=
\frac{\alpha k_2 \lambda_1^{*2}(k_1-2\alpha)}
{k_1\big[k_1k_2-\alpha\lambda_1^*\lambda_2(k_2-1)\big]}.
\end{equation}

This expression immediately implies $\bar{\delta}=0$ for $\alpha=0$, i.e., triadic transmission does not contribute at early times without nestedness.

\emph{Step 2: quasi-stationary ratio $\bar{\Pi}=\rho^{\rm SI}/\rho^{\rm I}$.}
We next impose $\dot{\rho}^{\rm SI}\simeq0$ and keep only leading terms in $\rho^{\rm I}$.
Near the disease-free state, we use $\rho^{\rm II}=\rho^{\rm I}-\rho^{\rm SI}=\mathcal{O}(\rho^{\rm I})$ and the pair closures
$\rho^{\rm SSI}\approx\rho^{\rm SS}\rho^{\rm SI}/\rho^{\rm S}\simeq\rho^{\rm SI}$ and
$\rho^{\rm ISI}\approx(\rho^{\rm SI})^2/\rho^{\rm S}=\mathcal{O}((\rho^{\rm I})^2)$,
so that terms involving $\rho^{\rm ISI}$ are negligible at leading order.
For the cross-order contributions, we express $\rho^{\rm ISI_\Delta}=\bar{\delta}\,\rho^{\rm SI}$ and retain only terms linear in $\rho^{\rm SI}$.
This yields a linear balance of the form
\[
0 \simeq -\mu\,\rho^{\rm SI}
+ \beta_1\Big[(k_1-1)\rho^{\rm SI}-\rho^{\rm SI}\Big]
+ \beta_2\,\frac{k_2}{k_1}\Big[(k_1-2\alpha)\rho^{\rm ISI_\Delta}-2\alpha\rho^{\rm ISI_\Delta}\Big]
+ \mu(\rho^{\rm II}-\rho^{\rm SI}),
\]
which, after substituting $\rho^{\rm II}=\rho^{\rm I}-\rho^{\rm SI}$ and
$\rho^{\rm ISI_\Delta}=\bar{\delta}\,\rho^{\rm SI}$, can be rearranged to obtain $\rho^{\rm SI}/\rho^{\rm I}$.
Evaluated at $\lambda_1=\lambda_1^*$, the resulting quasi-stationary value is
\begin{equation}
\label{eq:Pi_mf}
\bar{\Pi}\equiv\frac{\rho^{\rm SI}}{\rho^{\rm I}}
=
\frac{k_1}{2k_1-\lambda_1^*(k_1-2)+\lambda_2(4\alpha-k_1)\bar{\delta}},
\end{equation}
again matching the results shown in the main text.
Finally, since $\Psi\equiv\rho^{\rm ISI_\Delta}/\rho^{\rm I}=\delta\,\Pi$, we also have
$\bar{\Psi}=\bar{\delta}\,\bar{\Pi}$.

Together, $\bar{\Pi}$ and $\bar{\delta}$ quantify the early-time balance between dyadic and triadic contagion channels and provide a mechanistic decomposition of how nested hyperedges reshape the onset of spreading.

\begin{figure}
    \centering
    \includegraphics[width=\linewidth]{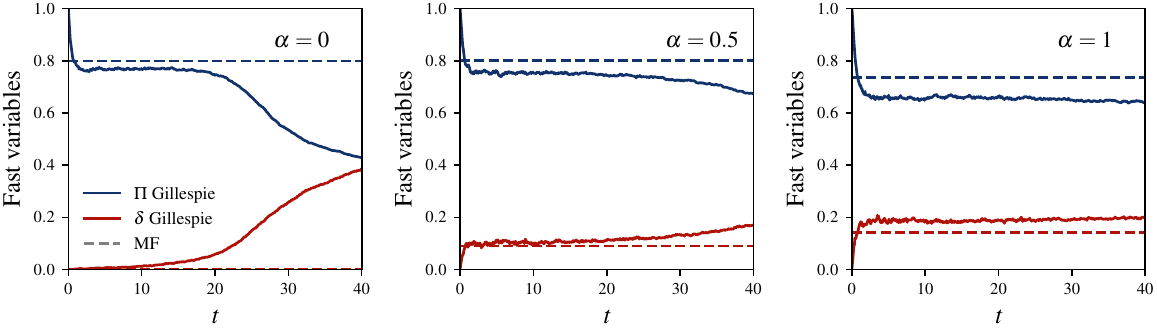}
\caption{\textbf{Quasi-stationary states of fast variables from temporal evolution.}
Temporal evolution of the fast variables $\Pi$ (blue) and $\delta$ (red) obtained from Gillespie simulations (solid lines) on regular random hypergraphs with $N=3000$, $k_1=5$, and $k_2=2$, shown for three representative values of the overlap $\alpha$. Simulations are performed at $\lambda_1=\lambda_1^*+0.05$ and $\lambda_2=3$. Dashed lines indicate the corresponding quasi-stationary values $\bar{\delta}$ and $\bar{\Pi}$  predicted by the analytical expressions in Eqs.~\eqref{eq:delta_mf} and~\eqref{eq:Pi_mf}.}

    \label{fig:sm_fastvariables}
\end{figure}

To validate these analytical quasi-stationary predictions, we compare them against direct Gillespie simulations, close to the epidemic threshold $\lambda_1^*$ for each value of $\alpha$, for $\lambda_2=3$.
Figure~S\ref{fig:sm_fastvariables} reports the temporal evolution of the fast variables $\Pi$ and $\delta$ obtained from Gillespie simulations, together with the corresponding quasi-stationary values $\bar{\Pi}$ and $\bar{\delta}$ predicted by Eqs.~\eqref{eq:Pi_mf} and~\eqref{eq:delta_mf}.  

After a short transient, both variables rapidly converge toward time-independent plateaus that are in good quantitative agreement with the analytical quasi-stationary predictions, for all values of the overlap $\alpha$ considered.  
Small deviations are expected and originate from the fact that the analytical expressions are evaluated exactly at the epidemic threshold $\lambda_1=\lambda_1^*$, while numerical simulations are necessarily performed in its vicinity. These discrepancies are further amplified by finite-size effects and stochastic fluctuations, which are particularly relevant since the corresponding densities are very small close to threshold.

Despite these limitations, the fast-variable approximation accurately captures the early-time organization of the dynamics.  
In particular, it correctly predicts that $\delta=0$ in the non-nested limit $\alpha=0$, reflecting the absence of group-mediated correlations, while $\delta$ becomes strictly positive as soon as $\alpha>0$, signaling the activation of group-embedded transmission pathways.

Overall, these results confirm that the ratios defining $\Pi$ and $\delta$ relax on a timescale well separated from the slow evolution of the global infected density, justifying their treatment as quasi-stationary quantities.  
The analytical expressions for $\bar{\Pi}$ and $\bar{\delta}$ therefore provide a robust microscopic characterization of how nested hyperedges reshape contagion pathways at the onset of spreading.

\end{document}